\documentstyle[psfig,amssymb,graphicx,epsf,latexsym,longtable,supertabular,rotating]{mn}     

\title[Optical and Infrared Spectroscopy of the type IIn SN 1998S :
Days 3-127]
{Optical and Infrared Spectroscopy of the type IIn SN 1998S :
Days 3-127}
\author[A.Fassia {\it et al.}]
{A. Fassia$^{1}$, W.P.S. Meikle$^{1}$,
N. Chugai$^{2}$, T.R. Geballe$^{3}$, P. Lundqvist$^{4}$,
N.A. Walton$^{5}$,   
\and D. Pollacco$^{6}$, S. Veilleux$^{7}$, G. Wright$^{8}$,
M. Pettini$^{9}$, T. Kerr$^{10}$, 
E. Puchnarewicz$^{11}$,  
\and P. Puxley$^{12}$,  M. Irwin$^{9}$,  C. Packham$^{13}$,
S. J. Smartt$^{9}$, D. Harmer$^{14}$  
\\ \\
$^1$Astrophysics Group, Blackett Laboratory, Imperial College, Prince
Consort Rd, London SW7 2BZ, UK\\
$^2$Institute of Astronomy, Russian Academy of Sciences, Ryatniskaya 48, 109017 Moscow, Russia\\
$^3$Gemini Observatory Northern Operations Center, 670 N. A'ohoku Place, Hilo,
Hawaii, 96720, USA  \\ 
$^4$Stockholm Observatory, SE-133 36, Saltj\"{o}baden, Sweden\\
$^5$Isaac Newton Group, Apartado de Correos 321, 38780 Santa
Cruz de La Palma, Islas Canarias, Spain \\
$^6$Astrophysics and Planetary Sciences Division, The Queen's University
of Belfast, Belfast BT7 1NN \\ 
$^7$Department of Astronomy, University of Maryland, College Park, MD
20742-2421\\ 
$^8$Royal Observatory, Blackford Hill, Edinburgh, EH9 3HJ, Scotland, UK \\
$^9$Institute of Astronomy, Madingley Road, Cambridge, CB3 0HA   \\
$^{10}$Joint Astronomy Centre, 660 N. A'Ohoku Place, University Park,
Hilo, Hawaii 96720, USA \\
$^{11}$Mullard Space Science Laboratory,University College, London,
Holmbury St. Mary, Dorking, Surrey RH5 6NT\\  
$^{12}$International Gemini Project, 950 N. Cherry Ave., P.O. Box
26732, Tucson, Arizona, 85726-6732, USA \\ 
$^{13}$Department of Astronomy, University of Florida, Gainesville, FL
32611 USA\\  
$^{14}$National Opt. Astr. Obs., P.O.Box 26732, Tucson, AZ 85726, USA\\
}
\begin{document}
\maketitle  
\begin{abstract}
We present contemporary infrared and optical spectroscopic
observations of the type~IIn SN~1998S covering the period between 3
and 127 days after discovery. During the first week the spectra are
characterised by prominent broad H, He and C~III/N~III emission lines
with narrow peaks superimposed on a very blue continuum
(T$\sim$24000~K). In the following two weeks the C~III/N~III emission
vanished, together with the broad emission components of the H and He
lines. Broad, blueshifted absorption components appeared in the
spectra.  The temperature of the continuum also dropped to
$\sim$14000~K. By the end of the first month the spectrum comprised
broad blueshifted absorptions in H, He, Si~II, Fe~II and Sc~II. By
day~44, broad emission components in H and He reappeared in the
spectra. These persisted to as late as $\sim$100-130 days, becoming
increasingly asymmetric.  We agree with Leonard {\it et al.} (2000)
that the broad emission lines indicate interaction between the ejecta
and circumstellar material (CSM) emitted by the progenitor.  We also
agree that the progenitor of SN 1998S appears to have gone through at
least two phases of mass loss, giving rise to two CSM zones.
Examination of the spectra indicates that the inner zone extended to
$\leq$ 90~AU, while the outer CSM extended from 185~AU to over
1800~AU.  

We also present high resolution spectra obtained at 17 and 36
days. These spectra exhibit narrow P~Cygni H~I and He~I lines
superimposed on shallower, broader absorption components. Narrow lines
of [N~II], [O~III], [Ne~III] and [Fe~III] are also seen.  We attribute
the narrow lines to recombination and heating following ionisation of
the outer CSM shell by the UV/X-ray flash at shock breakout.  Using
these lines we show that the outer CSM had a velocity of 40--50~km/s.
Assuming a constant velocity, we can infer that the outer CSM wind
commenced more than 170 years ago, and ceased about 20~years ago,
while the inner CSM wind may have commenced less than 9 years ago.
During the era of the outer CSM wind the outflow from the progenitor
was high - at least $\sim2\times 10^{-5}$M$_{\odot}$~yr$^{-1}$.  This
corresponds to a mass loss of at least $\sim$0.003~$M_{\odot}$,
suggesting a massive progenitor. The shallower, broader absorption is
of width $\sim$350~km/s and may have arisen from a component of the
outer CSM shell produced when the progenitor was going through a later
blue supergiant phase. Alternatively, it may have been produced by the
acceleration of the outer CSM by the radiation pressure of the UV
precursor.

We also describe and model first overtone emission in carbon monoxide
observed in SN~1998S.  We deduce a CO mass of $\sim$ 10$^{-3}$
M$_{\odot}$ moving at $\sim$2200~km/s, and infer a mixed metal/He core
of about 4~$M_{\odot}$, again indicating a massive progenitor.  Only
three core-collapse supernovae have been observed in the $K$-band at
post-100~days and all three have exhibited emission from CO.
\end{abstract}
\begin{keywords}  
 supernova, circumstellar matter
\end{keywords} 

\section{Introduction}
Core-collapse supernovae (SNe) are believed to arise from massive
progenitors (M$\geq$8-10M$_\odot$).  The lower-mass ($\sim$ 8-10
M$_{\odot}$) and higher-mass ($>20$ M$_{\odot}$) progenitors
experience heavy mass loss during the final stages of their evolution,
several solar masses being ejected via a range of mass-loss rates.
Consequently, in the vicinity of some core-collapse supernovae, dense
circumstellar material (CSM) would be distributed according to the
mass-loss history of the progenitor. The subsequent interaction of the
freely-expanding supernova ejecta with the slowly-moving CSM generates
a fast shock wave in the CSM and a reverse shock wave in the
ejecta. The shocked regions emit high-energy radiation.  The intensity
of this emission depends on the density of the CSM and the ejecta, and
the shock acceleration of the ejecta during the explosion (Chevalier
\& Fransson 1994). If the density of the CSM is small then the effects
of the interaction only become significant several years after the
explosion when the supernova has faded. However, if the CSM near the
supernova is relatively dense, strong CSM-ejecta interaction can begin
shortly after the explosion.  This is the case for type IIn (n=narrow
line) supernovae. These events exhibit narrow emission lines in their
spectra superimposed on broader emission profiles.  They also exhibit
a strong blue continuum (Schlegel 1990). However, the broad P~Cygni
absorption components typical of normal type~II SNe are weak or absent
in type~IIn SNe.  The presence of variable, narrow line emission is a
direct manifestation of the excitation of the dense CSM by the SN
radiation. In addition, the presence of broad H$\alpha$ emission
without a broad P Cygni absorption component is a clear indication
that the observed broad line emission is powered by the ejecta-wind
interaction (Chugai 1990).  Consequently, type~IIn supernovae can
provide unique information about the progenitor and its later
evolution, through the observed properties of their CSM. In addition,
the study of the interaction of these supernovae with the CSM can
provide vital clues about galaxy evolution and the nature of active
galactic nuclei ({\it c.f.}  Terlevich {\it et al.}  1992).

SN~1998S is the brightest type~IIn event ever observed.  It was
discovered on 1998 March 2.68 UT in the highly-inclined Sc galaxy NGC
3877 by Z. Wan (Li \& Wan 1998) at a broadband (unfiltered) optical
magnitude of +15.2.  The supernova is located at 16$^{\prime\prime}$
west and 46$^{\prime\prime}$ south of the nucleus.  By March 18.4 it
had brightened to $V=+12.2$ (Fassia {\it et al.} 2000). In a
prediscovery frame obtained on 1998 February 23.7 there was no
evidence of the supernova, down to a limiting apparent magnitude of
$\sim$+18 (Leonard {\it et al.} 2000, IAUC 6835).  We can thus assume
that SN~1998S was discovered within a few days of the shock breakout.
In the present paper we have adopted the discovery date 1998 March
2.68 UT = JD 2450875.2 as epoch 0 days, t$_{0}$, and express  all other epochs
relative to this fiducial date.  

Optical spectra obtained about day~3 (March~5--6) by Filippenko \&
Moran (1998), Huchra (Garnavich {\it et al.} 1998) and ourselves (see
$+$3.3 d spectrum in Figure~\ref{figop}) showed prominent H and He
emission lines with narrow peaks and broad wings superimposed on a
blue continuum.  As mentioned above these narrow lines indicate the
presence of a dense CSM in the vicinity of the supernova.  Using
optical spectropolarimetry obtained at 5~days, Leonard {\it et al.}
(2000) [L00] deduced that the CSM is asymmetrically distributed.
Bowen {\it et al.} (2000) have presented high resolution spectroscopy
of interstellar and circumstellar lines towards SN~1998S.  They
suggest that the CSM comprises a dense shell expanding at ~50~km/s
with a more highly ionised shell moving at $\sim$300~km/s.  They also
estimate that SN~1998S is at a real distance of $\sim$10~kpc from the
nucleus and deduce from the interstellar lines that it lies on the far
side of the galaxy disk.  Gerardy {\it et al.}  (2000) [G00] presented
near-IR spectra of SN~1998S spanning 95--355~days post-maximum light.
They identified emission from carbon monoxide. They also suggested
that late-time multi-peak H and He line profiles in their optical and
IR spectra indicate emission from a disk-shaped or ring-shaped
circumstellar component.  In addition, their t$\geq$225~d
near-infrared spectra exhibit a continuum that rises towards the
longer wavelengths. They propose that the rising continuum is likely
due to dust heated by the interaction of the ejecta with the CSM.

In Fassia {\it et al.} (2000) we presented contemporaneous optical
and infrared photometric observations of SN~1998S covering the period
between 11 and 146 days after discovery.  Using the interstellar
Na~I~D lines we derived an extinction $A_{V}=0.68^{+0.34}_{-0.25}$
mag.  We also examined the evolution of the total luminosity and found
that during the first month the luminosity decreased very rapidly {\it
viz.} $\sim$0.08~mag/day.  Subsequently (30-70 days) the decline rate
decreased, resembling that of type~IIL supernovae {\it viz.} 0.05
mag/day. By day~$\sim$100 the decline rate slowed to 0.01 mag/day
matching the radioactive decay of $^{56}$Co.  From the bolometric
luminosity after $\sim$100 we estimate that 0.15$\pm$0.05~M$_{\odot}$
of $^{56}$Ni were produced in the explosion.  Furthermore, we
discovered that as early as day 130 the supernova exhibited an
astonishingly high infrared (IR) excess, $K-L'=+2.5$.  We argue that
this excess is due to dust grains in the vicinity of the supernova.
However, the physical extent of this early IR luminosity source was so
large that the emission must have come from pre-existing dust in the
CSM, possibly powered by X-rays from the ejecta-CSM interaction.

In this paper we present optical and infrared spectroscopy of SN~1998S
covering the period 3--127~days after discovery.  The observations and
the reduction procedure are presented in section~2.  In section~3 we
discuss possible line identifications. In section~4 we present an
overview of the spectral behaviour of SN1998S and derive constraints
about the nature and the characteristics of the CSM. We also analyse
and model the first overtone of the CO emission. The work is
summarised in section~5.

\section{Observations}

\subsection{Optical spectroscopy}
\subsubsection{Low resolution spectroscopy} 
\begin{table*}
\centering
\caption{Log of optical spectroscopy of SN1998S}
\begin{minipage}{\linewidth}
\renewcommand{\thefootnote}{\thempfootnote}
\renewcommand{\tabcolsep}{0.13cm}
\begin{tabular}{ccclccccl} 
\\ 
JD \footnote{2450000+} 
& Date      & Epoch  & Telescope/ & Spectral & Spectral & Slit Width & Spectrophotometric  & Observer \\ 
& (1998 UT) &  (d)   & Instrument  & Range    & Resolution &  (arcsec)  &Standard\footnote{Used to correct for atmospheric and
instrumental transmission and to flux calibrate the supernova spectra.}               &          \\
& &  & & $\lambda\lambda$ (\AA) &  $\Delta\lambda$ (\AA) &  & \\
\\ 
878.5 & March 6.0  & 3.3  & INT/IDS    & 3850-10337 & 6.75  & 1.46 & SP1121+216  &  M. Irwin \\
889.5 & March 16.8 & 14.3 & INT/IDS    & 3390-9593  & 6.65  & 1.19 &
Feige 34 & C. Packham \\
&&&&&& & &\& D. Pollacco \\
892.6 & March 20.1 & 17.4 & WHT/UES    & 4190-8856  & 0.12  & 2.00  & -&J. Wood \\ 
&&&&&& & &\& S. Catalan \\
892.9 & March 20.4 & 17.7 & WIYN/Hydra & 3904-7094  & 6.45  & (3.1)\footnote{This is a fibre-fed spectrograph. The diameter of the fiber used was 3.1 arcsec.}
& GD140 & D. Willmarth \\
902.4 & March 29.9 & 27.2 & INT/IDS    & 3393-8899  & 10.05 & 1.59 & Feige 34 & D. Pollacco \\
911.5 & April 8.0  & 36.3 & WHT/UES    & 3620-9424  & 0.14  & 2.36 & -
& S. J. Smartt \\
913.7 & April 10.2 & 38.5 & WIYN/Hydra & 6425-6686  & 0.27  & (3.1)$^{c}$   & a$^{2}$CVn & D. Harmer \\
947.5 & May 14.0   & 72.3 & INT/IDS    & 3485-9602  & 8.2   & 1.38 & BD262606  & D. Pollacco \\
972.4 & June 7.9   & 97.2 & INT/IDS    & 3398-9230  & 10.04 & 1.19 & Feige 34 & D. Pollacco \\
\\ 
\end{tabular} 
\end{minipage}
\label{opspec}
\end{table*}

\begin{figure*}
\vspace{12.9cm}
\includegraphics{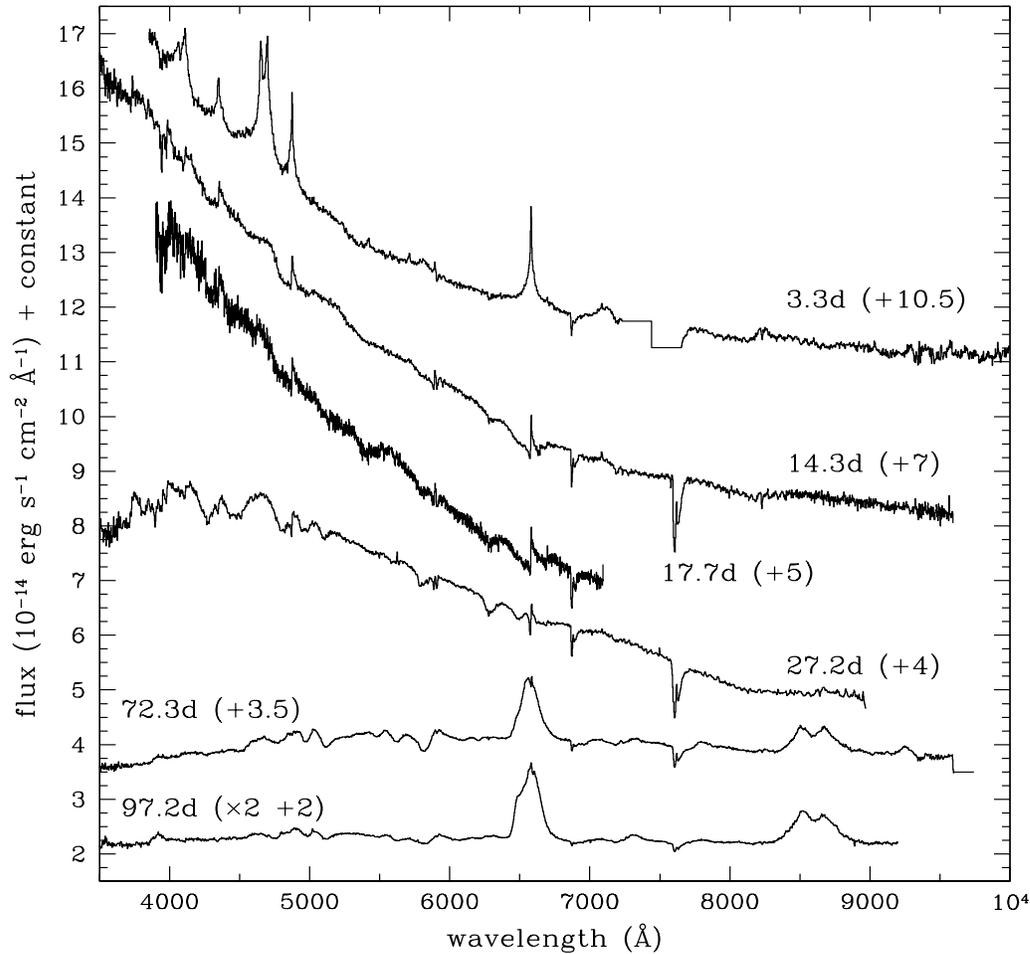}
\caption[]{Optical spectra of SN~1998S taken at the Isaac Newton
Telescope (La Palma) and the Wisconsin-Indiana-Yale-NOAO Telescope
(Kitt Peak) (see Table~\ref{opspec} for details).  The data have not been
corrected for redshift or reddening. The epochs are with respect to the
discovery date (1998 March 2.68 UT= 0~days).  For clarity, the spectra
have been displaced vertically by the amounts indicated (in units of
10$^{-14}$ ergs s$^{-1}$ cm$^{-2}$ \AA$^{-1}$).  The 97.2d
spectrum has also been multiplied by 2. }
\label{figop}
\end{figure*}
\begin{figure*}
\vspace{22.5cm}
\includegraphics{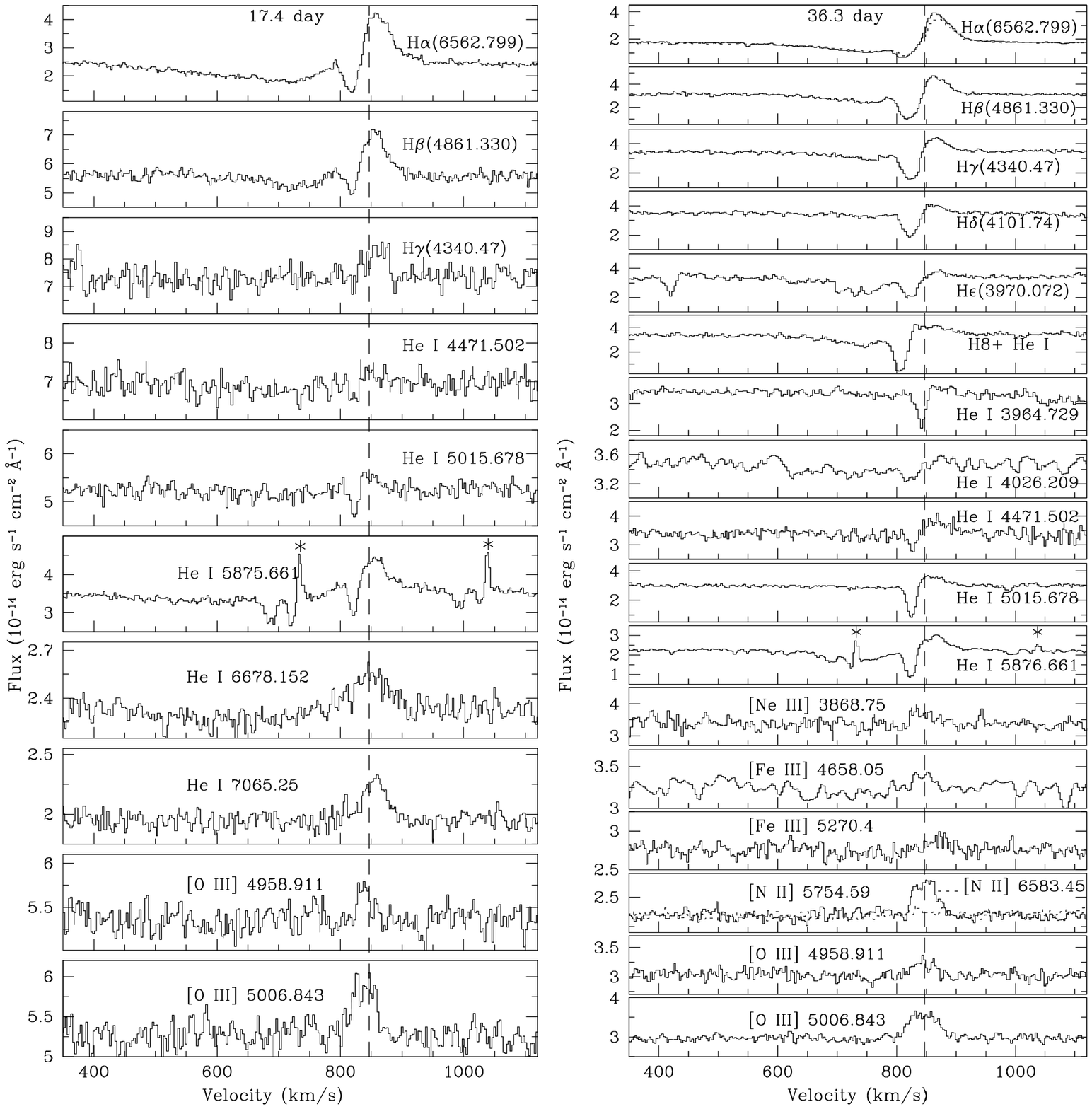}
\caption[]{High-resolution (echelle) optical spectra of SN~1998S taken
at the William Herschel Telescope (WHT) on days 17.4 and 36.3. Also
shown as a dotted line on the top-right plot is the high-resolution
spectrum of the $H_\alpha$ line obtained at the
Wisconsin-Indiana-Yale-NOAO (WIYN) Telescope on day 38.5(see
Table~\ref{opspec} for details).  The spectra have not been corrected
for redshift.  The Na~I $\lambda$5889, 5895 night sky lines are marked
with an asterisk.  The vertical dashed line corresponds to the adopted
redshift for the SN centre of mass, 847$\pm$3~km/s (cf. Section 3.2)}
\label{fighires}
\end{figure*}

Optical spectroscopy of SN 1998S was acquired using the ISIS
spectrograph and the UES echelle spectrograph at the William Herschel
Telescope (WHT), La Palma, the IDS spectrograph at the Isaac Newton
Telescope (INT), La Palma and the fiber-fed HYDRA spectrograph at the
Wisconsin-Indiana-Yale-NOAO Telescope (WIYN), Kitt Peak.  The log of
optical observations is shown in Table~\ref{opspec}.

Apart from the 17.4, 36.3d UES spectra, the spectroscopic data were
reduced by means of standard routines in the data-reduction package
FIGARO (Shortridge 1995). The data were first cleaned of cosmic
rays. Subsequently, the frames were debiased, flat-fielded and
sky-subtracted.  The spectra were extracted from the resulting frames
using simple extraction. Wavelength calibration was performed by means
of Cu-Ar and Cu-Ne lamps. The spectrophotometric flux standards used
are listed in Table~\ref{opspec}.

Owing to adverse weather conditions and the use of fairly narrow slits
(slit width $\leq$ 1.6$^{\prime\prime}$) in most cases, the fluxing
described above was not accurate. Therefore to improve the fluxing, we
made use of our $BVR$ photometry of SN~1998S (Fassia {\it et al.}
2000) to derive correction scaling factors.  Transmission functions
for the B, V and R bands were constructed using the JKT filter
functions, the CCD response and the standard La Palma atmospheric
transmission.  The spectra were then multiplied by these functions and
the resulting total flux within each band compared with the observed
magnitudes.  For dates where simultaneous photometric observations did
not exist we interpolated within the photometric data (Fassia {\it et
al.} 2000).  However, we do not have any photometry prior to 11~days.
Therefore, to scale the 3.3~d spectrum, we used $BVR$ photometry
(day~5) from Garnavich {\it et al.} (1999) together with $V$-band and
unfiltered CCD photometry from the IAU circulars (IAUC~6829, IAUC~6831,
IAUC~6835) covering days 0-7.5 to interpolate to day 3.3.  For any
particular epoch the scaling factors derived for each band agreed to
within $\pm$10\% demonstrating good internal consistency for the
procedure.  The spectra were multiplied by the geometric mean of the
scaling factors for each epoch.  The amounts by which the spectra had
to be scaled to match the photometry ranged from $\times$1.23 to
$\times$2.48.  Including uncertainties in the photometry, we estimate
the final absolute fluxing of the spectra to be accurate to $\le$15\%.
The optical spectra are shown in Figure~\ref{figop}.

\subsubsection{High resolution spectroscopy}

High resolution optical spectra were obtained on days~17.4 and 36.3
with the WHT UES echelle spectrograph, and on day~38.5 with the WIYN
Hydra spectrograph using a high dispersion grating.  Some of the UES
spectra have already been presented and discussed in Bowen {\it et
al.}  (2000).  The WHT/UES high-resolution spectra were reduced using
the IRAF package ECHELLE. The data were bias-subtracted and flat-field
corrected. The different echelle orders were then traced and extracted
from the frames. Wavelength calibration was by means of a
Thorium-Argon lamp.  We checked the precision of this calibration
using night-sky emission lines present in our spectra together with
the high-resolution night-sky line atlas of Osterbrock {\it et al.}
(1996).  This demonstrated that our wavelength calibration is accurate
to $\pm2$~km/s.

Cosmic rays were present in the extracted spectra.  These were
identified and removed by comparing repeat observations of the spectra
obtained at different times during the night.  The counts in the
contaminated pixels were then replaced by those of the corresponding
pixels in the other spectrum.  The mean fluxes of the spectra were
then normalised to unity by fitting a low-order spline function to the
instrumental continuum.  Finally the spectra were fluxed as follows.
First we determined the shape and flux of the overall SN continuum in
the optical region at days 17.4 and 36.3. This was done by fitting a
low-order spline function to the continuum in contemporary
low-resolution spectra.  This was straightforward for the earlier
epoch since a low-resolution spectrum was available on day~17.7.
However, for day~36.3 no contemporary low-resolution spectrum was
available.  We therefore used the 27.2~d low-resolution spectrum
scaled to match the $BVR$ magnitudes on day 36.3d. Given the evolution
of the $B-V$ and $V-R$ colours, we can argue that the continuum shape
did not change much during this period.  We then scaled the normalised
high-resolution spectra so that the continuum in the regions where
lines were observed matched the continua determined from the
low-resolution spectra.  We estimate the fluxing of the high
resolution lines to be accurate to better than $\pm$20\%.  In
Figure~\ref{fighires} we show the spectral regions where lines are
observed.  Also shown in Figure~\ref{fighires} is the high-resolution
38.5d WIYN/HYDRA spectrum.  The HYDRA spectra were reduced in the same
way as the low-resolution optical spectra (cf Section~2.1) and were
fluxed following the same procedure as for the day~36.3
high-resolution spectra.

\subsection{Infrared Spectroscopy}
Infrared spectra of SN~1998S were obtained at the United Kingdom
Infrared Telescope (UKIRT), Hawaii, using the Cooled Grating
Spectrograph CGS4.  The observing log is shown in
Table~\ref{irspec}.  During the observations the telescope was nodded
7.5~arc seconds along the slit.  The data were reduced using CGS4DR
(Daly \& Beard 1992) and FIGARO.  After bias-subtraction and
flat-field correction the pair of images obtained at the two nod
positions were subtracted to remove sky line emission.  If residual
sky line emission persisted, it was removed using POLYSKY.  The
positive and negative spectra were then extracted from the resulting
frames using the optimal extraction algorithm of Horne (1986).
Wavelength calibration was by means of argon and krypton arc
lamps. Observations of standard stars were used to correct for the
atmospheric and instrumental transmission and to flux-calibrate the
spectra.  Hydrogen absorption lines in the standard star
spectra were removed before these corrections/calibrations were
performed. The standard stars used are listed in Table~\ref{irspec}.

As with some of the optical spectra, variable weather conditions
together with the use of narrow slits for some of the spectra meant
that fluxing using standard stars was approximate only.  To improve
the fluxing we could, in principle, follow a similar procedure to that
applied to the optical spectra.  However, no IR photometry was
available until 13.8~d, 11~days after our earliest IR spectrum.
Moreover, for most of the $J$ band spectra the $J$ band photometry
transmission function (filter+atmosphere+detector response) extended
significantly beyond the red limit of the spectra.

For the earliest IR spectrum, which was at 2.8~d in $K$, there is a
particular problem in checking the fluxing.  Not only is there a lack
of accurate photometry around this time, but also optical observations
indicate that the light curve was rising rapidly, making it very
difficult to make a flux correction using extrapolated magnitude
values.  However, the spectrum was taken in a wide (2.46'') slit and
so fluxing by means of the standard star is accurate to
$\pm$20\%.  Therefore, no attempt was made to improve this fluxing.
\begin{table*}
\centering
\caption{Log of infrared spectroscopy of SN1998S obtained at UKIRT with CGS4}
\begin{minipage}{\linewidth}
\renewcommand{\thefootnote}{\thempfootnote}
\renewcommand{\tabcolsep}{0.13cm}
\begin{tabular}{cccccccl} \\ \\
JD \footnote{2450000+} 
& Date & Epoch &  Spectral Range  & Spectral  & Slit Width & Spectrophotometric & Observer \\ 
 &(1998 UT)& (d) & $\lambda\lambda$ (\AA) &  Resolution  & (arcsec)
&Standard \footnote{Used to correct for atmospheric and 
instrumental transmission and to flux calibrate the supernova spectra}& \\
&&&&$\Delta\lambda$ (\AA)   &&&\\
\\
878.0 & March 5.5  & 2.8   & 18870-25200 & 66.5 & 2.46 & HD84800 &  S. Veilleux \\
880.0 & March 7.5  & 4.8   & 10218-13405 & 32.5 & 2.46 & HD84800 &   S. Veilleux \\
882.0 & March 9.5  & 6.8   & 9700-12610  & 25.0 & 1.23 & HD84800 & E. Puchnarewicz \\
884.1 & March 11.6 & 8.9   & 18864-25232 & 57.5 & 1.23 & BS4761  & P. Puxley \\
887.1 & March 14.6 & 11.9  & 9926-13099  & 25.1 & 2.46 & BS4761  & G. Wright \\
888.0 & March 15.5 & 12.8  & 18834-25209 & 67.5 & 2.46 & HD105601 &  G. Wright \\
892.9 & March 20.4 & 17.7  & 10215-13402 & 20.5 & 1.23 & BS4431 & T. Geballe \\
919.8 & April 16.3 & 44.6  & 9921-13098  & 25.2 & 2.46 & BS4550 & T. Geballe \\
919.8 & April 16.3 & 44.6  & 14076-19739 & 49.0 & 2.46 & BS4550 & T. Geballe \\
919.8 & April 16.3 & 44.6  & 18857-25240 & 50.1 & 2.46 & HD105601 & T. Geballe \\
919.8 & April 16.3 & 44.6  & 10569-11117 & 4.3  & 2.46 & BS4550 & T. Geballe \\
930.7 & April 27.2 & 55.5  & 9770-12971  & 18.9 & 1.23 & BS4388 & M. Pettini \\ 
930.8 & April 27.3 & 55.6  & 10579-11129 & 3.3  & 1.23 & BS4388 & M. Pettini \\ 
983.8 & June 19.3  & 108.6 & 18747-25071 & 49.3 & 2.46 & BS4550 & T. Kerr \\
988.8 & June 24.3  & 113.6 & 10087-13285 & 16.9 & 1.23 & BS4761 & T. Geballe \\
1001.8 & July 7.3  & 126.6 & 14452-20478 & 37.6 & 1.23 & BS4572 & T. Geballe \\ \\
\end{tabular} 
\end{minipage}
\label{irspec}
\end{table*}

\begin{figure*}
\vspace{12cm}
\includegraphics{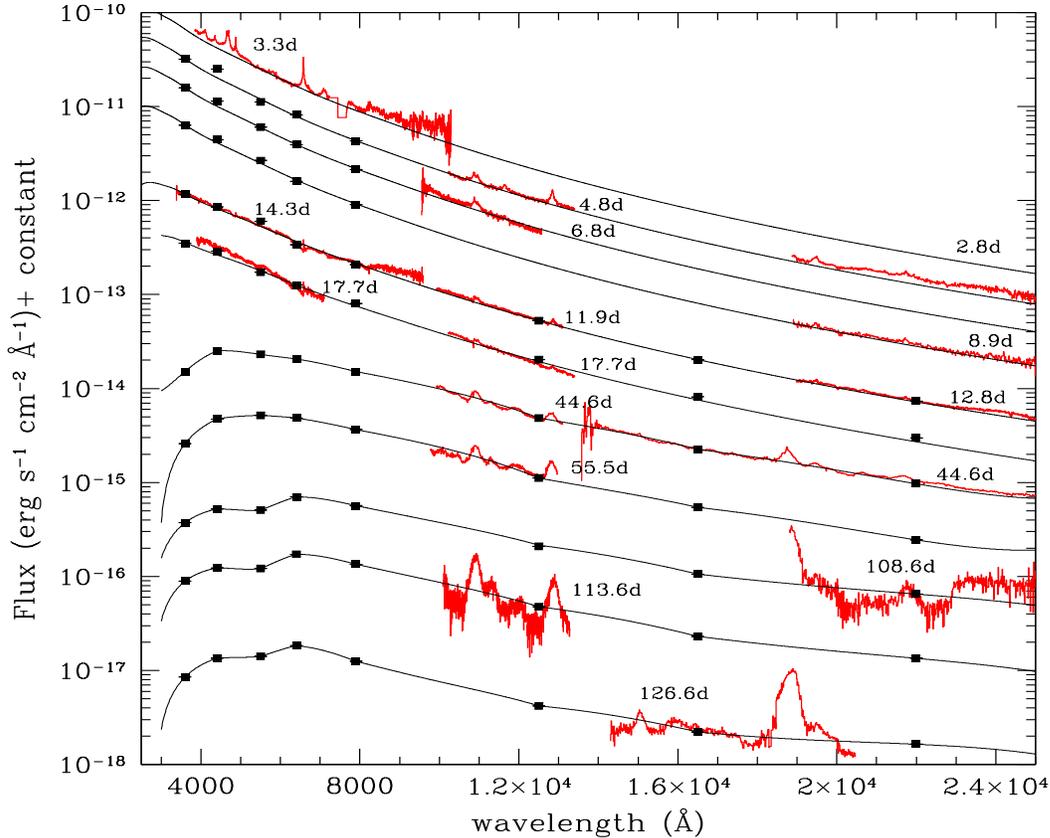}
\caption[]{Blackbody fits ($t<$44.6~days) and spline fits
($t\geq$44.6~days) to $UBVRIJHK$ photometry (see section 2.2 for
details). These fits were used to ascertain the absolute fluxing of
the infrared spectra.}
\label{blacb_ircal}
\end{figure*}

\begin{figure*}
\vspace{10.30cm}
\includegraphics{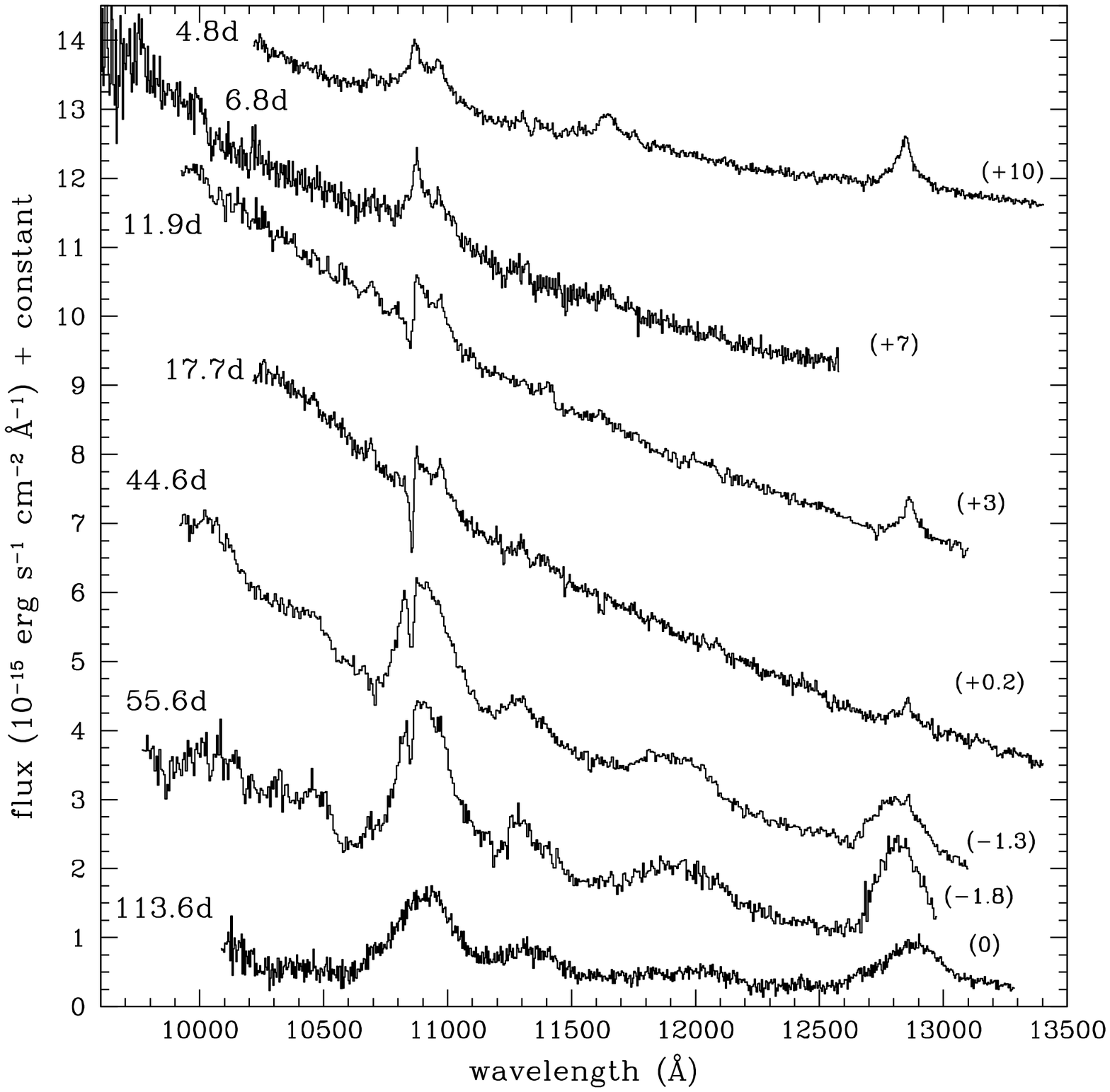}
\caption[]{$J$ band spectra of SN~1998S taken at the United Kingdom
Infrared Telescope (UKIRT).  The spectra have not been corrected for
redshift. The epochs are with respect to the discovery date (1998
March 2.68 UT= 0~days).  For clarity, the spectra have been displaced
vertically by the amounts indicated (in units of 10$^{-15}$ ergs
s$^{-1}$ cm$^{-2}$ \AA$^{-1}$). }
\label{figj}
\end{figure*}
\begin{figure*}
\vspace{10.30cm}
\includegraphics{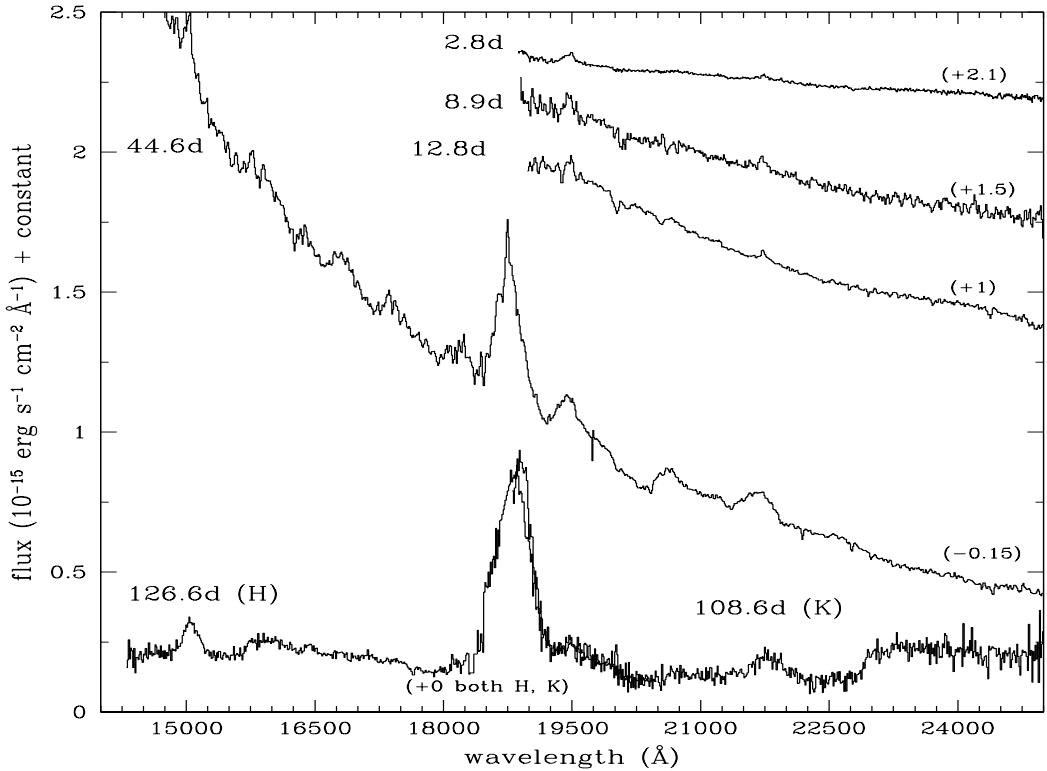}
\caption[]{$H$ and $K$ band spectra of SN~1998S taken at the United Kingdom
Infrared Telescope (UKIRT).  The spectra have not been
corrected for redshift. The epochs are with respect to the
discovery date (1998 March 2.68 UT= 0~days).  For clarity, the spectra
have been displaced vertically by the amounts indicated (in units of
10$^{-15}$ ergs s$^{-1}$ cm$^{-2}$ \AA$^{-1}$). }
\label{fighk}
\end{figure*}

For the other IR spectra we corrected the fluxing as follows.  Using
the optical and IR magnitudes given in Fassia {\it et al.} (2000),
together with earlier optical magnitudes reported by Garnavich {\it et
al.}  (1999) we interpolated to provide magnitudes for each epoch upon
which IR spectra were acquired.  Thus, a set of $UBVRI$ magnitudes
were established for epochs 4.8~d to 8.9~d, and $UBVRIJHK$ magnitudes
for epoch 11.9~d onward.  The magnitudes were then converted to fluxes
using the calibrations of Wilson {\it et al.}  (1972) and Bessel
(1979).  For epochs 4.8~d to 8.9~d we fitted reddened blackbody
functions to the $UBVRI$ data. The blackbody functions were reddened
using the empirical formula of Cardelli {\it et al.} (1989) with
$A_{V}=0.68^{+0.34}_{-0.25}$ (Fassia {\it et al.} 2000).  We then
scaled the fluxes of the infrared spectra so that the continua matched
the blackbody functions (cf. Figure~\ref{blacb_ircal}).  This
procedure ignores the contribution of the optical emission lines to
the blackbody fits, probably leading to an overestimate of the IR
flux.  However, from the 3.3~d optical spectrum, we estimate this
systematic error was no more than 7\%.  For epochs 11.9~d to 17.7~d,
while contemporaneous IR photometry was available, correction of the
$J$-band spectra was still difficult due to the incomplete overlap
problem explained above.  Therefore, for the 11.9 and 17.7~d $J$-band
spectra we followed a similar procedure to that carried out for the
earlier epochs, except that now we included $JHK$ photometry in the
blackbody fits.  Given the slow variation of the supernova flux around
this time ({\it cf.} Fassia {\it et al.} 2000, Garnavich {\it et
al.}), we simply scaled the 11.9~d $J$ spectrum so that its continuum
matched the reddened blackbody 14.3~d optical spectrum. The
12.8~d 18834--25240~\AA\ spectrum was also flux-corrected by matching
it to the same reddened blackbody.  This spectrum was also checked
using contemporaneous $H$ and $K$ light curves (see below).
Similarly, the 17.7~d $J$-band spectrum was matched to the 17.7~d
optical spectrum.  

By epoch 44.6~d and later, the IR spectra were increasingly affected
by emission/absorption features (cf. Figures~\ref{figop}, ~\ref{figj},
~\ref{fighk}).  Consequently, flux correction using the blackbody
fitting procedure was increasingly inappropriate.  Instead, for these
later epochs we applied spline fits (Press {\it et al.} 1992) to the
$UBVRIJHK$ photometry.  We then scaled the infrared spectra so that
the integrated fluxes of the observed IR spectra and spline spectra
were equal in the wavelength regions where they coincided
(cf. Figures~\ref{blacb_ircal}).   Scaling factors for the infrared spectra
ranged from $\times$0.92 to $\times$1.65.

To check the above scaling procedures we used the optical flux
correction method (Section~2.1) for the few IR cases where this was
possible.  Thus, using contemporary IR photometry (Fassia {\it et al.}
2000), we estimated absolute flux levels for the 12.8~d and 108.6d $K$
spectra and the 126.6~d $H$ spectrum.  We constructed transmission
functions for the $H$ and $K$ bands using the IRTF/UKIRT filter
functions and the standard Mauna Kea atmospheric transmission.  We
then multiplied the corresponding spectra by these functions and
compared the resulting total fluxes with the photometry.  The scaling
factors derived by this procedure agreed to within 12\% with those
derived by the blackbody/spline fits mentioned above. We estimate,
therefore, that the absolute fluxing of the corrected infrared spectra
is accurate to $\lesssim$15\%.  The final fluxed infrared spectra are
shown in Figures~\ref{figj} and ~\ref{fighk}.

\begin{figure}
\vspace{7.5cm}
\includegraphics{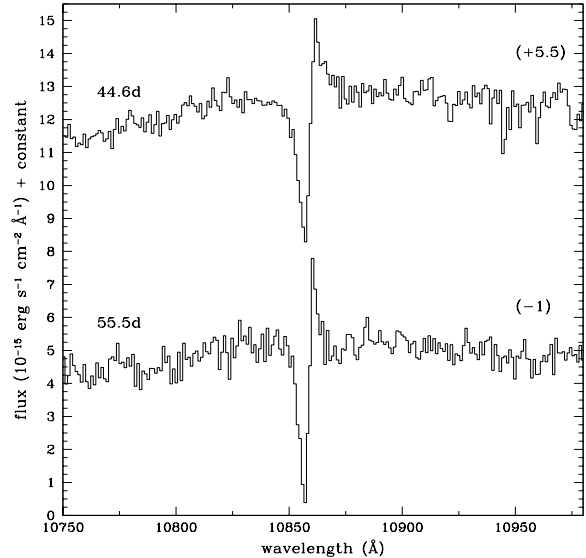}
\caption[]{High-resolution spectra of SN~1998S taken at the United
Kingdom Infrared Telescope (Hawaii).  The epochs are with respect to
the discovery date (1998 March 2.68 UT= 0~days).  For clarity, the
spectra have been displaced vertically by the amounts indicated (in
units of 10$^{-15}$ ergs s$^{-1}$ cm$^{-2}$ \AA$^{-1}$).}
\label{figirhi}
\end{figure}

On days~44.6 and 55.6, high resolution ($\sim$100~km/s) spectra of the
He~I 10830~\AA\ line were acquired with UKIRT/CGS4.  They were
wavelength calibrated using sky lines from the raw images.  Fluxing
was carried out by scaling their continua to match the contemporary
low-resolution spectra.  They are shown in Figure~\ref{figirhi}.

\section{Description of the Spectra}
\subsection{Low-Resolution Spectra}
In Figures~\ref{opspec_ncont}, \ref{irspec_ncont} we redisplay the
optical and IR spectra, respectively, but with the best
blackbody/spline fit continua (see Section~2.2) subtracted in order to
minimise distortion of the line profiles.  In addition the relative
flux is plotted logarithmically in order to bring out the detail in
the line profiles.  The spectra have not been redshift corrected.  The
resulting line identifications are indicated in Tables~3, 4 and in
Figures~\ref{opspec_ncont}, \ref{irspec_ncont}.  In general, the lines
have complex profiles comprising mixtures of broad and narrow emission
and absorption lines.  To examine these profiles in detail we carried
out multiple component fits using Gaussian profiles.  This provided
peak wavelengths, intensities and widths of the separate components of
each profile. These are listed in Tables~3, 4 for the optical and
infrared spectra respectively.  We note that the presentations of L00
and G00 include first season optical and IR spectra which have similar
epochs to some of the spectra discussed here.

\subsubsection{3--7~days}
\begin{figure*}
\vspace{10.7cm}
\includegraphics{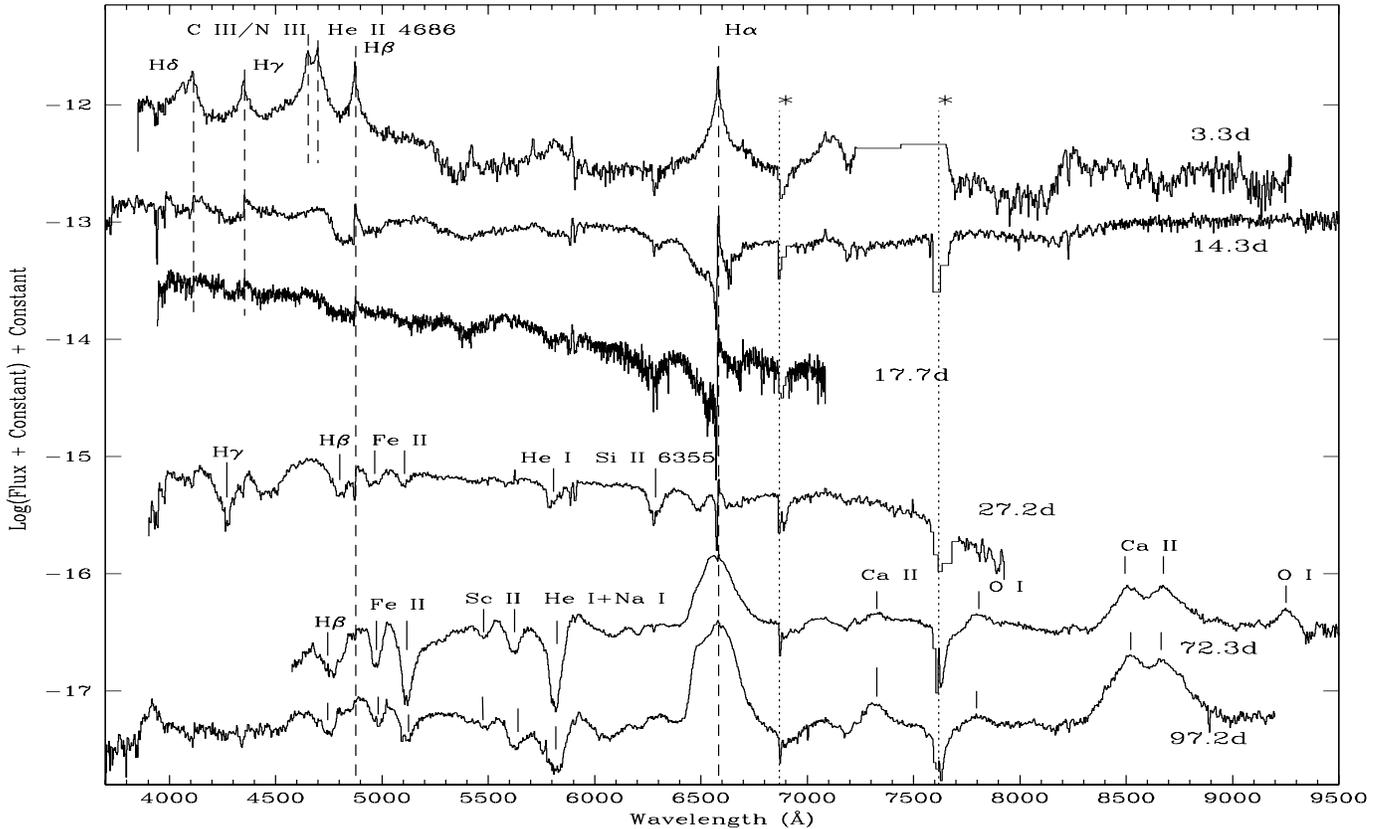}
\caption[]{Optical spectra of SN 1998S with blackbody/spline fit
continua shown in Figure~\ref{blacb_ircal}
subtracted. Suggested line identifications are indicated.(The dashed
lines correspond to the rest wavelength of the lines in NGC 3877). Residual
telluric absorption lines are marked with the dotted lines and an asterisk.}
\label{opspec_ncont}
\end{figure*}

\begin{figure*}
\vspace{10.7cm}
\includegraphics{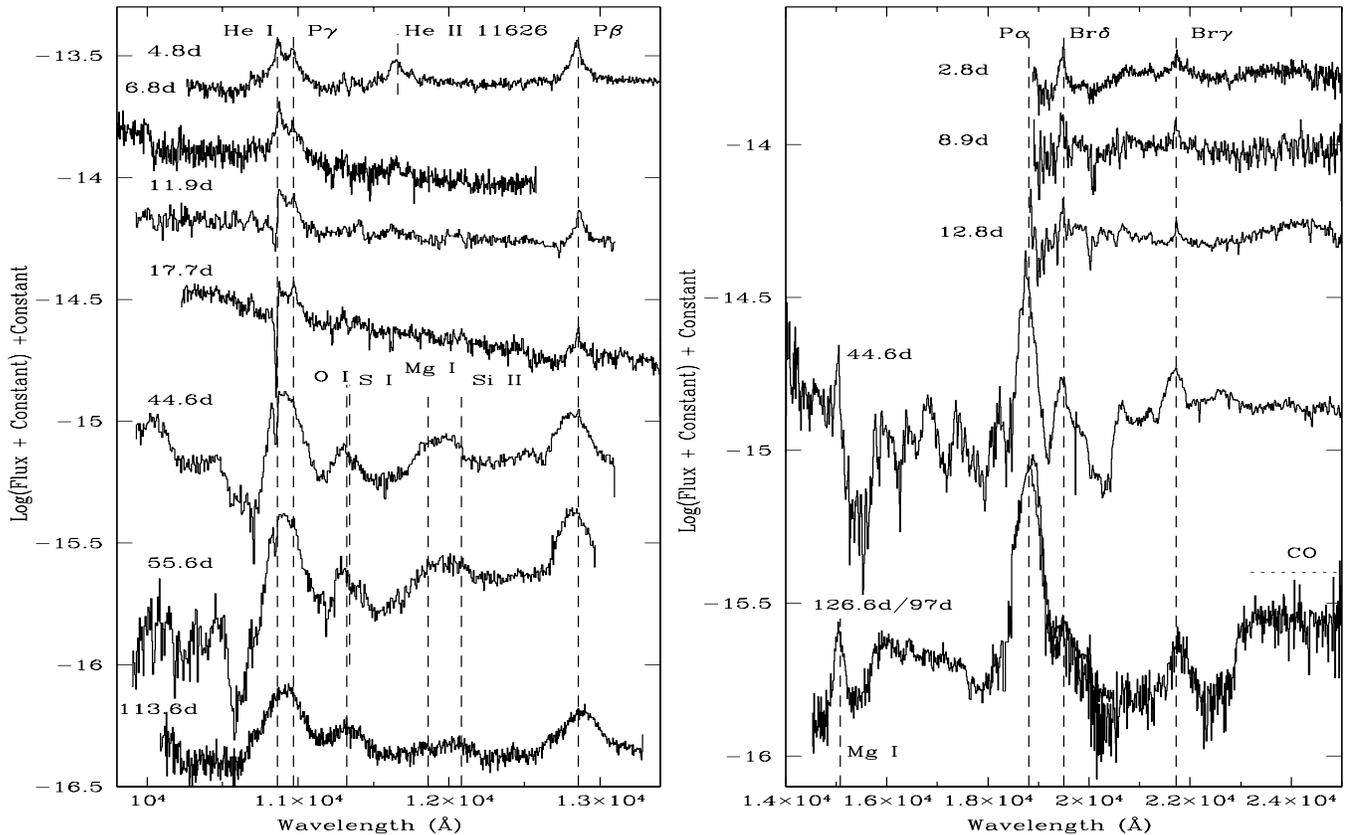}
\caption[]{Infrared spectra of SN 1998S with the blackbody/spline fit
continua shown in Figure~\ref{blacb_ircal} subtracted. Suggested line
identifications are indicated.(The dashed lines correspond to the rest
wavelength of the lines in NGC 3877) }
\label{irspec_ncont}
\end{figure*}

During the first week, the spectra are characterised by strong,
complex emission lines of the Balmer, Paschen and Brackett series,
He~I~10830~\AA, He~II 4686~\AA\ and C~III~4648/N~III~4640~\AA, lying
on a smooth blue continuum.  There is little sign of any absorption
components.  In Figure~\ref{blacb_ircal} we show blackbody spectra
reddened by E(B-V)=0.22 (R$_V$=3.1) superimposed on the optical and IR
continua.  The blackbody temperature on day~3.3 is 23,600$\pm$500~K.
A somewhat higher blackbody temperature of 28,000~K is derived by L00
for their day~4 optical spectrum, using a slightly higher E(B-V) of
0.23.  They point out that this is quite high for an SN~II at this
epoch.  In agreement with L00, we also find that the day~3--4
blackbody fit fails to account for all the continuum blueward of
$\sim$5000~\AA.  L00 argue that the most likely cause is a departure
from the blackbody spectrum due to the increase of the continuum
absorptive opacity (bound-free, free-free) with wavelength.  We
confirm the L00 finding that a typical line profile at this time
comprises a narrow unresolved component superimposed on a broad base
({\it cf.} Figures~\ref{opspec_ncont}, \ref{irspec_ncont}).  This is
illustrated in Figures~\ref{figha_vel} and ~\ref{fighe_vel} where we
show multiple gaussian component fits to H$\alpha$ and He~I~10830~\AA\
respectively.  The unresolved component has a FWHM $\leq$ 500 km/s,
superimposed on a broad component of FWHM $\sim$4000~km/s.  Other
broad emission features include one at $\sim$7100~\AA\ which we
attribute to a blend of He~I 7065~\AA\ and C~II~7100~\AA.  Other
narrow lines include C~III 5696~\AA, He~I 5876~\AA, 6678~\AA,
7065~\AA\ and He~II 5411~\AA, 8236~\AA\ (possibly) and 11626~\AA.
Note that the narrow absorption feature at 20,000~\AA\ is residual 
absorption due to H$_{2}$O, CO$_{2}$ and CH$_{4}$ vapour in the
earth's atmosphere.


\begin{figure*}
\vspace{0.95\textheight} 
\includegraphics{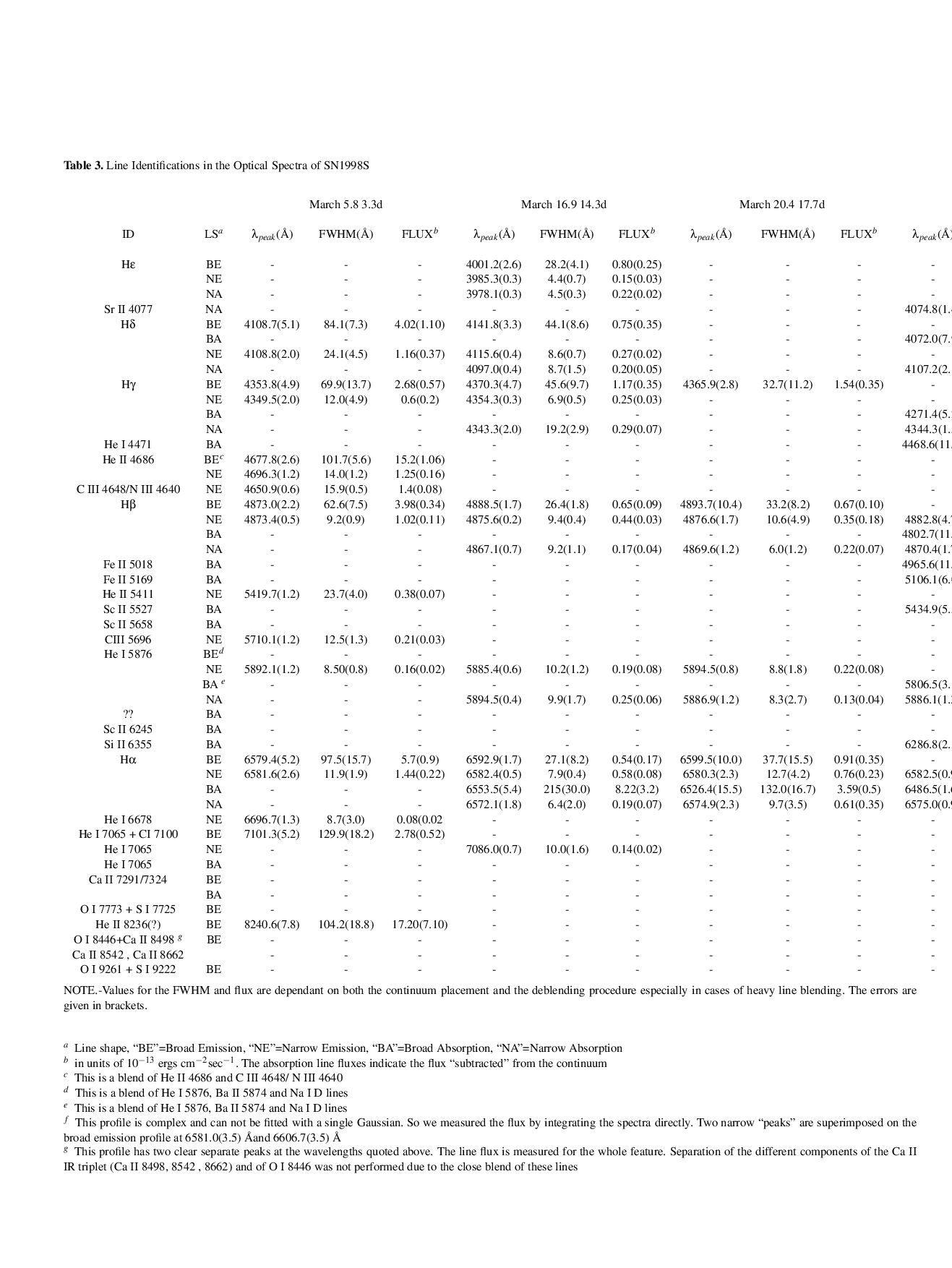}
\end{figure*}
\addtocounter{table}{1}

\subsubsection{8--18~days}
Between days~3.3 and 17.7, blackbody fits indicate a cooling
continuum.  By day~17.7 the blackbody temperature was 14500~K,
consistent with the 13,400$\pm$1200~K derived by Fassia {\it et al.}
(2000) from contemporary photometry.  During this time, the lines also
underwent a substantial change.  Between days~3.3 and 14.3 the broad
emission component in Balmer, Paschen, Brackett and He~I 10830~\AA\
lines largely vanished (see also Qiu {\it et al.} 1998), leaving
behind narrower, but still quite complex, emission features.  The
intense C~III~4648/N~III~4640~\AA\ feature disappeared altogether.
The disappearance of the broad emission in the Balmer and He~I
10830~\AA\ lines was accompanied by the appearance of a broad,
blueshifted absorption extending to velocities of $\sim$10000km/s.
However, broad absorptions did not appear in the Paschen or Brackett
lines.  In H$\alpha$ and He~I 10830~\AA\ weak, unresolved (FWHM $\leq$
500 km/s) P~Cygni profiles developed.  These narrow lines are examined
in greater detail in section~3.2.

\subsubsection{27--56~days}
The day~27 optical spectrum is very similar in appearance to the
day~25 spectrum of L00.  In making line identifications, we were
guided by those given for the photospheric spectra of the type II SN
1987A and SN 1995V (Williams 1987, Fassia {\it et al.} 1998).  By
day~27 the spectrum comprises broad, blueshifted absorptions in H
(Balmer series), He~I+Na~I blend, Si~II, Fe~II and Sc~II, together
with narrow P~Cygni lines of H (Balmer series), He~I and narrow
emission lines of [O~III].  (Narrow emission lines of [N~II], [Ne~III]
and [Fe~III] were also seen around this time in our high resolution
spectra - see section 3.2).  The FWHM of the broad absorptions is
$\sim$4000 km/s, with the minima blueshifted by 4000--5500km/s with
respect to the local standard of rest.  The blue edge of the H$\alpha$
absorption trough is blueshifted by $\sim$8000 km/s.  In the IR
spectra, by days 44 and 55, broad emission features
(FWHM~$\sim$5000km/s) had developed in the Paschen, Brackett and He~I
10830~\AA\ lines. The FWHM of the Pa$\alpha$, Pa$\beta$ and Br$\gamma$
lines on day~44.6 is $\sim$5000~km/s. A broad, blueshifted absorption
in He~I 10830~\AA\ is also apparent.  The deep narrow absorption seen
in He~I 10830~\AA\ on day~17 is still present on days 44 and 55. This
is discussed in section 3.2.  Other broad IR features that had
appeared by this period we identify with blends of
O~I~11287/94--S~I~11306~\AA, and Mg~I 11828--Si~II~12047~\AA\ ({\it
cf.}  Meikle {\it et al.} 1989).  L00 note that the relative strengths
of some of the broad features in the optical spectrum resemble those
found in the spectra of SNe~Ic, suggesting hydrogen-deficient ejecta.
They also suggest that the low contrast of the broad lines
relative to the continuum may be due to external illumination of the
line-forming region by light from the CSM-ejecta interaction (see also
Lentz et al. 2000 and Branch et al. 2000). This is discussed in more
detail in Section~4.1.2.

\subsubsection{72-127~days}
Between days 27 and 72 in the optical region the most striking change
is the replacement of the fairly weak, broad P~Cygni H$\alpha$ feature
with a strong, broad (FWHM $\sim$7000 km/s) emission feature ({\it
cf.} Figure~\ref{opspec_ncont}). The profile is somewhat asymmetric.
Its peak is blueshifted by $\sim$--500km/s with respect to the SN rest
frame, and its blue wing is steeper. The extreme limit of the blue
wing is blueshifted by $\sim$--8500 km/s.  The observation of
comparably wide Paschen emission lines on day~44.6 suggests that broad
H$\alpha$ emission had probably already appeared by then.  In other
Balmer lines, the change from days 27 to 72 was less dramatic.  The
absolute depth of the broad absorption in H$\beta$ weakened and the
H$\gamma$ absorption disappeared almost entirely by day~97. A dramatic
change between days~27 and 72 is the appearance of strong, broad
emission in the Ca~II~IR triplet.  The presence of O~I~7773~\AA\ and
O~I~9264~\AA\ emission lines on the 72~d spectrum suggests that
significant emission from the O~I~8446~\AA\ line may be blended with
the Ca~II multiplet.  The broad absorption lines in the He~I+Na~I
blend, Fe~II and Sc~II became more pronounced, but vanished in Si~II
6355~\AA.  The narrow H~I and He~I lines had weakened by day~72, with
only H$\alpha$ being clearly visible. Weak narrow [O~III]4959,
5007~\AA\ lines are visible in the day~97 spectrum.

Our day~97 optical spectrum is similar to the day~108 spectrum of L00,
and our day~109/113/127 composite IR spectrum is similar to the
day~110 spectrum described in G00.  By this period, the spectra were
dominated by broad emission lines of H$\alpha$, Pa$\alpha$, Pa$\beta$,
Pa$\gamma$+He~I 10830~\AA\ blend and the Ca~II triplet.  The
H$\alpha$, Pa$\alpha$ and Pa$\beta$ lines had broadened to a FWHM of
7000--8500~km/s.  Moreover the H$\alpha$, and possibly Pa$\alpha$,
profiles had become more asymmetric, with the development of a
pronounced steep blue edge. The asymmetry, and the pronounced red
wings of these lines could indicate that Thomson scattering of line
radiation occurred in the ejecta during this epoch.  We note however
that Br$\gamma$ had weakened significantly.  Also of importance is the
appearance by day~109 of first overtone emission from carbon monoxide.
This is discussed further in section~4.2.  L00 suggest that the
appearance of broad, asymmetric H$\alpha$ emission in their day~108
spectrum is due to interaction of the ejecta with that part of the CSM
responsible for the narrow lines seen at earlier epochs. We support
this proposal, and suggest that the presence of strong, broad Paschen
and Brackett lines on day~44 indicates that the interaction may have
already begun as early as this.  This will be discussed further in
Section 4.

\begin{table}
\begin{minipage}{\linewidth}
\caption{Line identifications for the infrared spectra
of SN 1998S.} 
\footnotesize
\label{ir_ida}
\renewcommand{\tabcolsep}{0.06cm}
\begin{tabular}{lllllll}
 DAY  &   ID     &   $\lambda_{peak}$(\AA)  &   $\lambda_{peak}$(\AA)
 & LS$^{a}$ & FWHM   &
\hspace*{0.3cm}FLUX$^{b}$ \\
&        	 &  (Emis.)       &  (Absorp.)      &  & (\AA)  & \\
\\
2.8   & Br$_\delta$               &  19470(20)  &   & BE   &220(55)  &  0.85(0.25) \\
      & Br$_\delta$               &  19494(6)   &   & NE   & 29(8)   &  0.1(0.05) \\ 
      & Br$_\gamma$               &  21775(35)  &   & BE   & 380(90) &  0.3(0.1) \\  
      & Br$_\gamma$               &  21735(6)   &   & NE   & 45(15)  &  0.06(0.03) \\
      &                           &             &   &      &         &             \\
 4.8  & He I 10830+P$_\gamma$     &  10910(15)  &   & BE   &241(35)  &  13.9(3.1) \\
      & P$_\gamma$                &  10966(8)   &   & NE   &32 (8)   &  0.8(0.3) \\
      & He I 10830                &  10869(8)   &   & NE   &32 (8)   &  1.3(0.2) \\
      & He II 11626?              &  11653(10)  &   & BE   &110(25)  &  4.0(0.8) \\
      & P$_\beta$                 &  12846(20)  &   & BE   &153(42)  &  6.1(1.5)  \\
      & P$_\beta$                 &  12847(8)   &   & NE   &38 (8)   &  1.4(0.4)  \\
      &                           &             &   &      &         &             \\
 6.8  & He I 10830+P$_\gamma$     &  10930(20)  &   & BE   &210(22)  & 14.9(1.9) \\
      & He I 10830                &  10875(5)   &   & NE   &29 (4)   & 1.8(0.3) \\
      &                           &             &   &      &         &         \\
 8.9  & Br$_\delta$               &  19469(20)  &   & BE   &133(40)  & 1.1(0.3) \\  
      & Br$_\gamma$               &  21726(20)  &   & BE   &115(25)  & 0.5(0.1) \\
      &                           &             &   &      &         &         \\
 11.9 & He I 10830                & & 10850(5)  &     NA   &26(7)    & 2.1(0.6) \\
      & He I 10830                & 10874(5)    &   & NE   &50(10)   & 3.1(0.9) \\
      & He I 10830+P$_\gamma$     & 10950(20)   &   & BE   &126(20)  & 6.9(0.6) \\ 
      & P$_\gamma$                & 10973(4)    &   & NE   &17(5)    & 0.4(0.2)  \\
      & P$_\beta$                 & 12868(10)   &   & BE   &101(20)  & 3.2(0.5)  \\ 
      & P$_\beta$                 & 12864(4)    &   & NE   &26(7)    & 0.7(0.3) \\
      &                           &             &   &      &         &           \\
 12.8 & Br$_\delta$               & 19470(30)   &   & BE   &96(25)   &  0.5(0.2) \\
      & Br$_\delta$               & 19497(12)   &   & NE   &23(8)    & 0.10(0.04) \\
      & Br$_\gamma$               & 21770(25)   &   & BE   &225(65)  & 0.35(0.15) \\
      & Br$_\gamma$               & 21729(6)    &   & NE   &35(15)   & 0.13(0.04) \\
      &                           &             &   &      &         &           \\
 17.7 & He I 10830                &&  10856(4)  &     NA   &27(9)    & 2.1(0.8)  \\
      & P$_\gamma$                & 10975(4)    &   & NE   &23(9)    & 0.7(0.4)  \\
      & He I 10830                & 10872(2)    &   & NE   &25(6)    & 1.7(0.4) \\
      & He I 10830+P$_\gamma$     & 10945(20)   &   & BE   &170(40)  & 7.5(2.5) \\
      & P$_\beta$                 & 12855(5)    &   & NE   &21(8)    & 0.5(0.2) \\
      & P$_\beta$                 & 12858(25)   &   & BE   &150(50)  & 3.0(1.2) \\
      &                           &             &   &      &          &           \\
 44.6 & He I 10830+P$_\gamma$     && 10642(30)  &     BA   &226(35)  & 11.0(2.5)\\
      & He I 10830+P$_\gamma$     & 10915(30)   &   & BE   &210(35)  & 30.5(5.5) \\
      & He I 10830                &&  10852(5)  &     NA   &23(0.5)  & 0.25(0.08) \\
      & O I 11287/11294             & 11300(15)   &   & BE   &150(25)  & 5.8(0.7) \\
      & +S I 11306                &             &   &      &         &          \\
      & Si II 12047               & 11935(11)   &   & BE   &280(15)  & 14.5(1.9) \\
      & +Mg I 11828               &             &   &      &         &          \\
      & P$_\beta$                 & 12817(10)   &   & BE   &216(15)  & 17.9(1.5) \\
      & Mg I 15031                & 15027(15)   &   & BE   &105(30)  & 2.1(0.4) \\
      & O I 18243/18021           & 18196(26)   &   & BE   &300(45)  & 2.7(0.4) \\
      & P$_\alpha$                & 18763(20)   &   & BE   &341(30)  & 16.7(1.5) \\
      & P$_\alpha$                & 18749(4)    &   & NE   & 34(6)   & 0.65(0.15) \\
      & Br$_\delta$               & 19460(20)   &   & BE   &285(40)  & 3.5(0.6) \\ 
      & Br$_\gamma$               & 21690(25)   &   & BE   &335(40)  & 3.3(0.4) \\
      &                           &             &   &      &         &           \\
 55.5 & P$_\delta$                & 10045(10)   &   & BE   &205(35)  & 14.9(5.0) \\
      & P$_\delta$                &&10044(4)    &     NA   & 25(9)   & 1.7(0.6) \\
      & He I 10830+P$_\gamma$     &&10600(20)   &     BA   & 130(30) & 8.5(1.8) \\ 
      & He I 10830                &&10851(5)    &     NA   & 17(8)   & 1.4(0.5) \\
      & He I 10830+P$_\gamma$     & 10910(20)   &   & BE   &205(35)  & 41.2 (1.9) \\
      & O I 11287/11294           & 11300(10)   &   & BE   &165(35)  & 10.0(2.5) \\
      & +S I 11306                &             &   &      &         &          \\ 
      & Si II 12047               & 11950(20)   &   & BE   &345(40)  & 18.0(1.7) \\
      & +Mg I 11828               &             &   &      &         &          \\
      & P$_\beta$                 & 12825(15)   &   & BE   &205(25)  & 29.5(2.5) \\
      &                           &             &   &      &         &        \\
\end{tabular}
\end{minipage}
\end{table}

\normalsize
\begin{table}
\contcaption{}
\footnotesize
\begin{minipage}{\linewidth}
\renewcommand{\tabcolsep}{0.05cm}
\begin{tabular}{lllllll}
\\
 DAY  &   ID     &   $\lambda_{peak}$(\AA)   &   $\lambda_{peak}$(\AA)
 & LS\footnote{Line shape,
``BE''=Broad Emission, ``NE''=Narrow Emission, ``BA''=Broad Absorption,
``NA''=Narrow Absorption }& FWHM      &
\hspace*{0.2cm}FLUX\footnote{in units of 10$^{-14}$  ergs
cm$^{-2}$sec$^{-1}$. The absorption line fluxes indicate the flux
``subtracted'' from the continuum} \\ 
     &   	&  (Emis.)     &  (Absorp.)     &   &(\AA)  & \\
\\
108.6-& He I 10830+P$_\gamma$     &  10910(15)  & &BE     & 275(25)  & 35.0(7.0)\\
126.6 & O I 11287                 & 11327(15)   & &BE     & 250(30)  & 14.5(3.5) \\
      &    +S I 11306             &             & &       &          &          \\
      & Si II 12047               & 12017(37)   & &BE     & 330(55)  & 6.5(2.0) \\
      & +Mg I 11828               &             & &       &          &          \\
      & P$_\beta$                 & 12875(15)   & &BE     & 300(30)  & 20(2.5) \\
      & Mg I 15031                & 15053(20)   & &BE     & 208(35)  & 2.3(0.4) \\
      & Br series,                & BLEND       & &BE     &    --      & 9.8(2.5)\footnote{This is the
integrated flux from 15500-17700~\AA} \\
      & Si I 15888, [Fe II],        &           & &       &          &    \\
      & [Si I], Mg I              &             & &       &          &    \\
      & P$_\alpha$                & 18833(40)   & &BE     & 430(50)  & 33.5(6.5) \\
      & Ca I 19309-19961          & 19585(65)   & &BE     & 480(90)  & 6.7(2.5) \\
      & Br$_\gamma$               & 21770(45)   & &BE     & 490(85)  & 3.9(1.2) \\
      & CO                        & $\sim$23400 & &BE     & --            & $>$30 \\
\\
\end{tabular}
NOTE.-Values
for the FWHM and flux are dependant on both the continuum
placement and the deblending procedure especially in cases of heavy
line blending. The errors are given in brackets.
\end{minipage}
\end{table}
\normalsize

\begin{figure}
\vspace{10cm}
\includegraphics{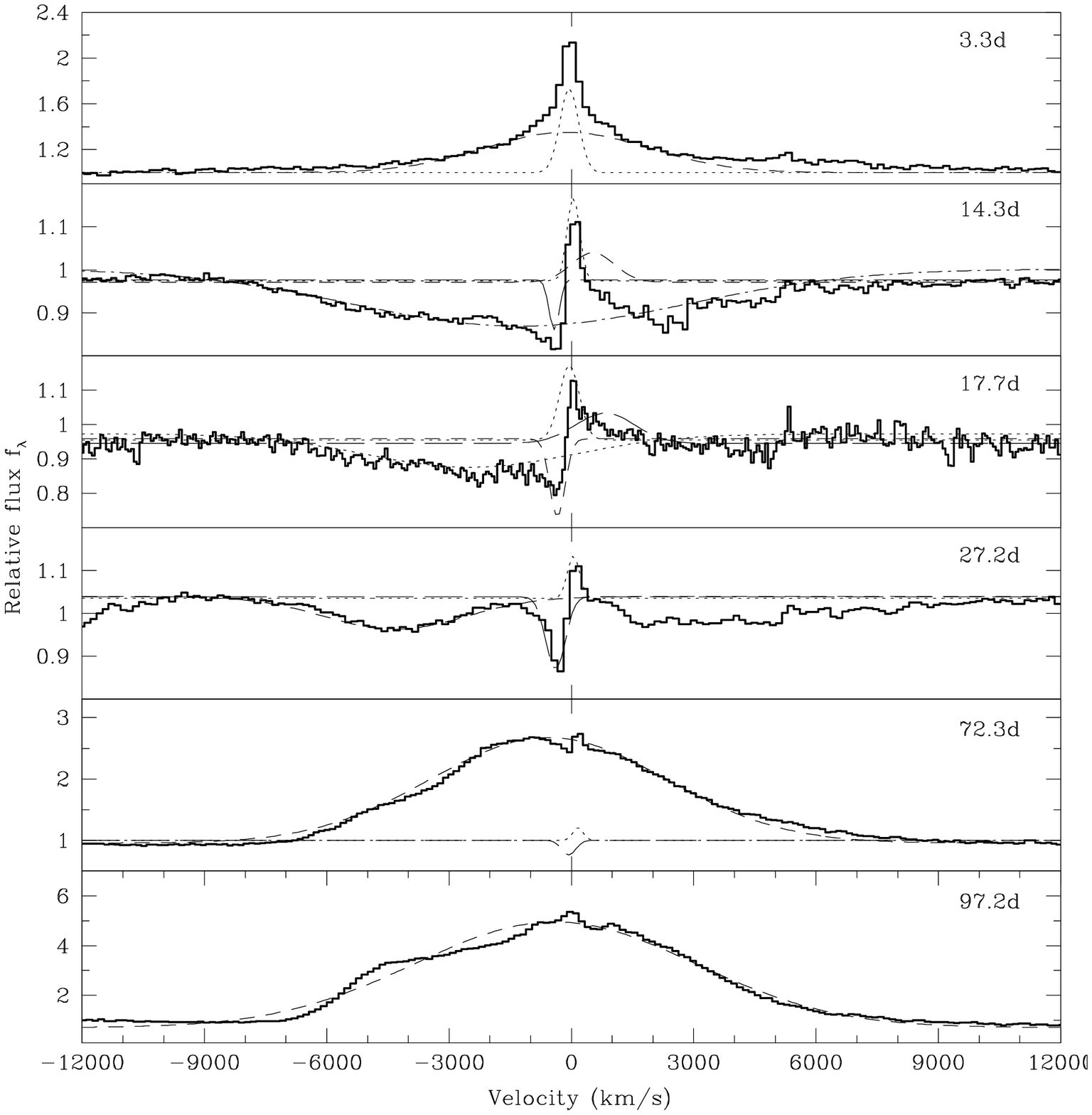}
\caption[]{Evolution of the H$\alpha$ line profile. Also shown are the
Gaussian components that were used to fit the observed line
profiles. The 0 km/s velocity corresponds to the adopted +847km/s
velocity for the SN center of mass (see Section~3.2)}
\label{figha_vel}
\end{figure}  

\begin{figure}
\vspace{10cm}
\includegraphics{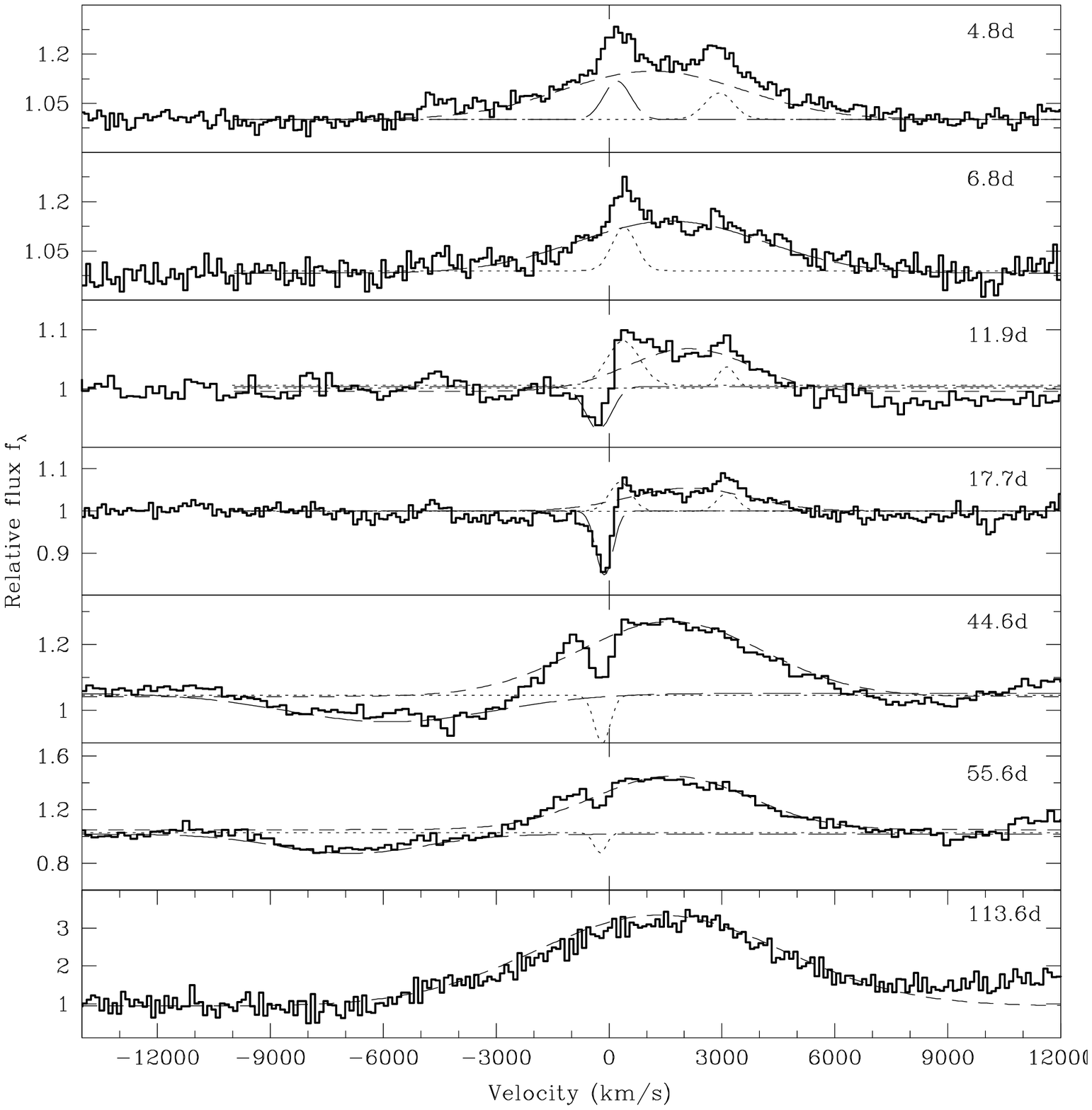}
\caption[]{Evolution of the He~I~10830 line profile. Also shown are
the Gaussian components that were used to fit the observed line
profiles. The 0 km/s velocity corresponds to the adopted +847km/s
velocity for the SN center of mass (see Section~3.2) }
\label{fighe_vel}
\end{figure} 

\subsection{High-Resolution Spectra}
The WHT/UES observations of SN~1998S on days~17.4 and 36.3 revealed
narrow absorption and emission lines.  Many of the absorption lines
have widths of around 2--15~km/s and can be attributed to the
interstellar medium (ISM) of the Milky Way and NGC~3877.  Fassia {\it
et al.} (2000) used the Na~I~D absorption lines observed on day~36.3
to estimate the extinction to SN~1998S.  In a more comprehensive
analysis of the WHT/UES ISM data, Bowen {\it et al.} (2000) examined a
range of lines in the UV and optical regions.  They also describe
a blueshifted absorption feature in Mg~II~2796~\AA\ and in the Balmer
lines, of width $\sim$350~km/s.  Variation in the depth of these
features between different epochs points to an origin in the CSM.  A
variable, narrow component in Na~I~D is also attributed to the CSM.
In addition, Bowen {\it et al.} report the presence of narrower
($\sim$30~km/s) Balmer and optical He~I P~Cygni lines as well as
similarly narrow P~Cygni lines in a number of Fe~II UV multiplets.
These too are attributed to the CSM of SN~1998S.  They also indicate
(in their Table~4) the presence on day~36.3 of narrow forbidden lines
of [N~II], [Ne~III], [O~III] and [Fe~III], and suggest an origin in
either the CSM or some other line-of-sight emission line region.

In the present work, we consider only those narrow lines whose origin
is in the CSM of SN~1998S.  In Figure~\ref{fighires} we display and
identify all the CSM lines present in the day~17.4 and 36.3 WHT/UES
spectra (resolution $\sim$7~km/s).  Also shown is the day~38 H$\alpha$
profile (resolution $\sim$12~km/s) obtained at the WIYN.  In
Figure~\ref{figirhi} we show the high resolution ($\sim$100~km/s)
P~Cygni spectra of the He~I 10830~\AA\ line acquired with UKIRT/CGS4
on days~44.6 and 55.6.  In the optical spectra we confirm the presence
of most of the allowed and forbidden lines described in Bowen {\it et
al.} (2000).  In particular, we confirm that the H~lines are composed
of two components {\it viz.} a narrow P~Cygni-like profile
superimposed on a broader absorption feature which extends as far as
$\sim$--350~km/s to the blue.  We also note the presence of a similar
broad absorption in He~I 5876~\AA\ and possibly, in He~I 4471,
5016~\AA.  Identifications of the narrow lines are listed in
Table~\ref{fighires_id} along with the redshifts of the emission
peaks, the blueshifts of the P~Cygni troughs relative to the estimated
SN centre of mass (see below), the widths of the emission and
absorption components and the line intensities.  The narrow H profiles
span a range of $\sim$130~km/s, the He~I profiles span $\sim$100~km/s,
while those of the forbidden lines span $\sim$80km/s.

Two of the narrow emission lines which Bowen {\it et al.} (2000) list in
their Table~4, they identify as He~I 3870~\AA\ and O~I 7773~\AA.  We
confirm the presence of these features.  However, it seems unlikely
that they originate in the SN CSM.  Firstly, both lines are
unresolved, implying a linewidth of less than $\sim$10~km/s.  This is
much narrower than the FWHM$\sim$60~km/s which is typical of the emission
components of other narrow lines.  The ``He~I 3870~\AA'' feature can
be seen in Figure~\ref{fighires} lying $\sim$100~km/s redward of
[Ne~III] 3869~\AA.  In the case of this line, comparison of two
successive repeat frames shows that it appeared strongly only in the
first, while in the second it was not discernible above the noise.  In
the case of the ``O~I 7773~\AA" line we note that it extends the full
length of the slit (15'') equivalent to over 1~kpc at NGC~3877.  It is
also at the wrong redshift for it to be associated with SN~1998S.  It
therefore seems unlikely that these two lines originate in the SN CSM.
We also note that in Table~4 of Bowen {\it et al.}, ``He~I~4290~\AA'' should
probably be read as ``He~I~4920~\AA''.  Bowen {\it et al.} also list a
narrow P~Cygni line identified with He~I~4713~\AA.  While there is a
weak feature at this wavelength in our reduced data, its S/N is very
low and so we do not include it in Table~\ref{fighires_id}.  Narrow
optical lines identified by us but not listed by Bowen {\it et al.} include
He~I 6678, 7065~\AA\ and [Fe~III]~4658~\AA.
\begin{figure}
\vspace{16cm}
\includegraphics{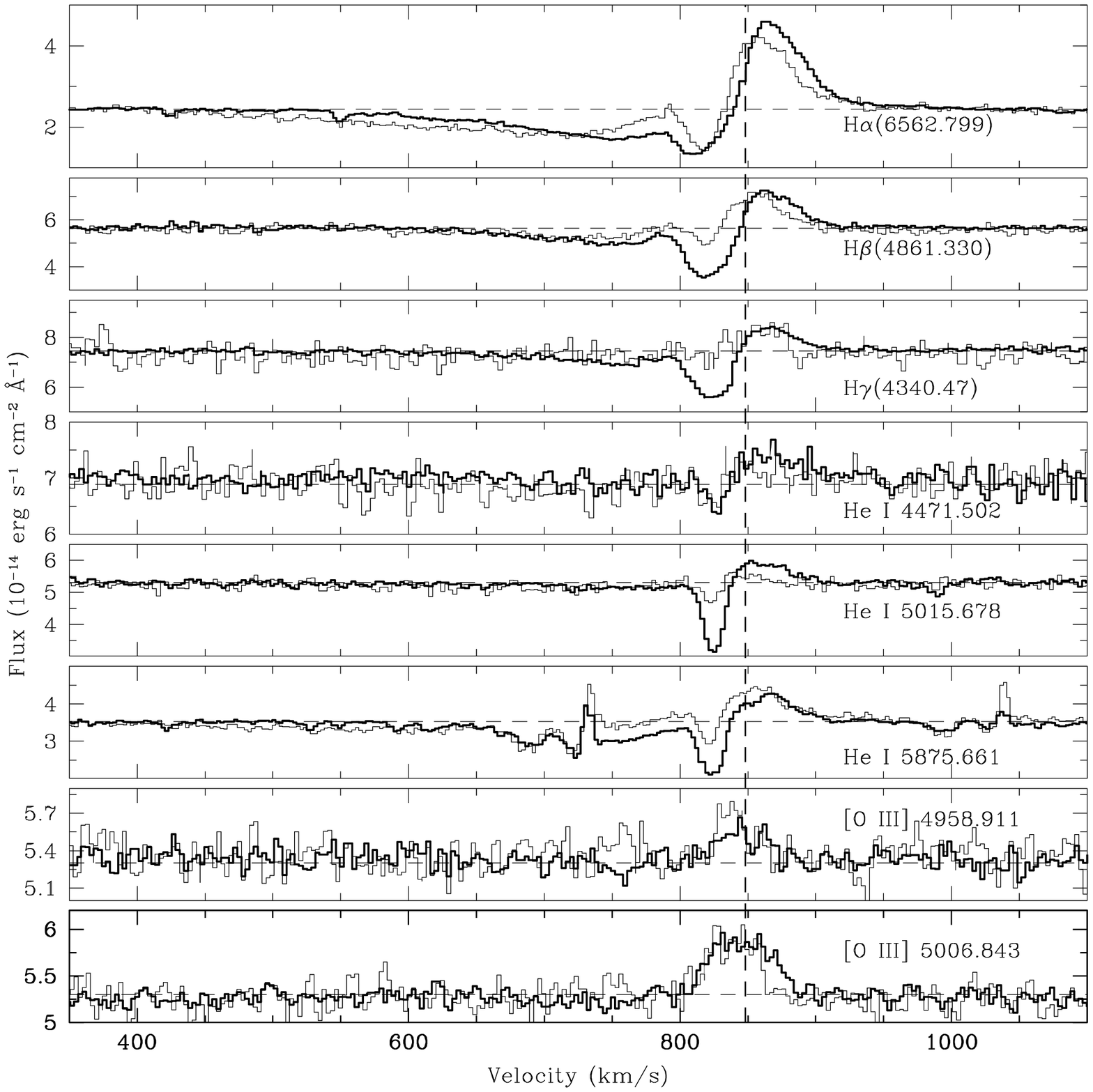}
\caption[]{Narrow line profiles observed in the 17.4d (fainter line)
and 36.3d (darker line) high-resolution spectra.  The vertical dashed
line corresponds to the adopted redshift for the SN centre of mass,
847$\pm$3~km/s (cf. Section 3.2).  To help the comparison of the line
profiles, the 36.3d continua have been displaced vertically to
coincide with the 17.4d continua.}
\label{fighivel}
\end{figure}

\begin{table*}
\caption{Narrow lines observed at high resolution. 
All observations were on days 17.4 or 36.3 unless indicated otherwise.}
\begin{minipage}{\linewidth}
\label{fighires_id}
\begin{tabular}{lllllll}
& \multicolumn{3}{c}{17.4 days} & \multicolumn{3}{c}{36.3 days} \\
\\
       ID\footnote{The rest wavelengths were obtained from the NIST
                   Atomic Spectra Database and the CRC Handbook of
                   Chemistry and Physics, 79th Edition.}

      & Shift\footnote{For emission lines, total redshift is given.
                       For absorption lines, shift is with respect to
                       SN centre of mass (+846.9~km/s).  Errors are
                       given in brackets}
      & FWHM  & Flux\footnote{in units of 10$^{-14}$ ergs
    cm$^{-2}$sec$^{-1}$. The absorption line fluxes indicate the flux
    ``subtracted'' from the continuum.  Errors are
                       given in brackets}
&Shift$^{b}$      & FWHM  & Flux$^{c}$\\
          &   (km/s) & (km/s)   &  & (km/s) & (km/s)&  \\ 
\\  

{\it Emission Lines} &  & \\
$ [Ne III]~3868.75$    & --       &  --    &    --      & +845(11)  &  55(12) & 0.40(0.10)  \\
 H8+He~I~3889.843\footnote{We adopted $\lambda$=3888.652~\AA\ for this blend}
	               & --       &  --    &    --      & +866(15)  & 108(25) & 1.67(0.82)\\
 He~I~3964.729         & --       &  --    &    --      & +852(15)  &  72(20) & 0.28(0.12)\\
 H$\epsilon$(3970.072) & --       &  --    &    --      & +867(11)  &  64(16) & 0.74(0.48)\\ 
 He~I~4026.209         & --       &  --    &    --      & +880(13)  &  37(11) & 0.07(0.03)\\
 H$\delta$(4101.74)    & --       &  --    &    --      & +859(7)   &  65(11) & 0.84(0.34)\\
 H$\gamma$(4340.47)    & 859(10) & 42(3)  & 0.79(0.15) & +862(7)   &  55(3)  & 0.89(0.03) \\
 $[O~III]$ 4363.209    & --       &  --    & $<$0.40    & +844(10)  &  65(12) & 0.26(0.07)  \\
 He~I 4471.502         & 851(18) & 60(33) & 0.22(0.10) & +865(8)   &  103(22)& 0.65(0.15) \\
 $[Fe~III]$ 4658.05    & --       &  --    & --    & +846(10)  &  39(16) & 0.14(0.05)\\
 H$\beta$(4861.330)    &  855(5) & 43(9)  & 0.96(0.21) & +862(6)   &  46(9) & 1.54(0.21) \\
 $ [O~III]$ 4958.911   &  840(5) & 30(7)  & 0.23(0.03) & +845(11)  &  48(7) & 0.22(0.05) \\
 $ [O~III]$ 5006.843   &  836(5) & 48(9)  & 0.65(0.12) & +847(7)   &  59(9) & 0.65(0.11) \\ 
 He~I 5015.678         &  847(6) & 39(13) & 0.19(0.08) & +858(7)   &  57(16)& 1.05(0.25) \\
 $[Fe~III]$~5270.40    & --       &  --    & --    & +868(14)  &  60(15)& 0.17(0.06) \\
$ [N~II]$ 5754.59      & --       &  --    & --    & +848(5)   &  45(3) & 0.30(0.04) \\
 He~I~5875.661         &  854(5) & 40(3)  & 0.52(0.07) & +863(5)   &  42(5)& 0.79(0.17) \\ 
 H$\alpha$ (6562.799)  &  855(5) & 55(5)  & 2.15(0.58) & +864(5)   &  55(7) &2.09(0.04) \\
 H$\alpha$$^{e}$(6562.799)& --   &   --    &   --       & +868(5)   &  50(5) &1.88(0.17)  \\
 $ [N~II]$ 6583.45     & --       &  --    &   --       & +843(7)   &  90(40)& 0.09(0.04) \\
 He~I~6678.152         & 851(5)  & 70(7)  & 0.44(0.13) & --         &  --   &   --     \\
 He~I~7065.25          & 855(5)  & 50(5)  & 0.43(0.04) & --         &  --   &   --   \\ 
 He~I~10830$^{f}$      & --       &  --   &  --   & +876(40)  &  $<$100 & 1.56(0.35)\\
 He~I~10830$^{g}$      & --       &  --   &  --   & +840(40)  &  $<$100 & 0.9(0.30) \\
&& \\
{\it Absorption lines} && \\
 H8+He I 3889.843       & --       &  --    &     --     & --41(7)  & 23(2) & 1.11(0.02)\\
 He~I~3964.729          & --       &  --    &     --     & --24(4)  & 16(3) & 0.31(0.02) \\
 H$\epsilon$(3970.072)  & --       &  --    &     --     & --23(5)  & 30(6) & 0.69(0.14) \\
 He~I~4026.209          & --       &  --    &     --     & --28(8)  & 27(3) & 0.08(0.02)  \\
 H$\delta$(4101.74)     & --       &  --    &     --     & --23(5)  & 30(13)& 0.90(0.08)\\
 H$\gamma$(4340.47)     & --       &  --    &     --     & --22(5) & 36(3)  & 1.13(0.07)\\
 He I 4471.502     	& --28(8)  & 13(10) & 0.79(0.10) & --23(10) & 25(10)& 0.27(0.05) \\
 H$\beta$(4861.33)      & --28(5)  & 25(6)  & 0.35(0.11) & --26(5)  & 42(4) & 1.78(0.25) \\
 He~I 5015.678     	& --25(4)  & 16(3)  & 0.21(0.05) & --21(4)  & 16(3) & 0.87(0.07)\\
 He~I~5875.661    	& --27(4)  & 15(3)  & 0.26(0.04) & --25(5)  & 19(3) & 0.61(0.04)\\
 H$\alpha$ (6562.799)   & --26(5)  & 32(5)  & 1.19(0.10) & --33(5)  & 45(14)& 1.06(0.44)\\
 H$\alpha$\footnote{Observed at 38.5~d} & -- &   -- & --  &--27(8)   & 40(10)& 0.87(0.07)\\
 He~I~10830\footnote{Observed at 44.6~d} & -- &   -- & --  & --100(60)  & $<$100& 2.28(0.55) \\
 He~I~10830\footnote{Observed at 55.6~d} & -- &   -- & --  & --100(60) & $<$100& 2.1(0.55) \\
\\
\end{tabular}

\end{minipage}
\end{table*}

The forbidden lines on day~36.3 all exhibit about the same absolute
redshift and width (see below), with a weighted mean redshift of
+846.9$\pm$2.9~km/s. We adopt this as probably being closest to the SN
centre of mass redshift, for the following reasons.  We note below
that between days~17.4 and 36.3 the red wings of the emission
components of the H lines (and possibly also the He~I lines) and
forbidden lines moved to the red by $\sim$+10--20~km/s.  We suggest
that this is a manifestation of light travel time effects across the
CSM around the supernova.  The radius of the outer CSM is at least 10
light days (see below) and so on the timescale of the observations as
the SN-flash paraboloid moves further into the CSM on the far side of
the SN, increasing amounts of more redshifted material will contribute
to the emission lines.  We therefore favour the later epoch
(36.3~days) since by that time the emission lines should be
originating from a greater proportion of the CSM {\it i.e.} the bias
to the blue due to light travel time effects should be reduced.  We
also favour the forbidden lines since the allowed lines appear to be
considerably more complex, comprising a mixture of broad and narrow
P~Cygni components plus a probable additional recombination emission
component.  Our adopted redshift for the supernova CSM, +847~km/s, is 
consistent with the approximate value of +860~km/s indicated in Bowen
{\it et al.} (2000).  We note that measurements of the systemic
velocity of NGC~3877 vary from +838$\pm$11~km/s using optical
observations of the nuclear region (Fouque {\it et al.} 1992), to
902$\pm$6~km/s (De Vaucouleurs {\it et al.} 1991) and +910$\pm$6~km/s
(Broeils \& van Woerden 1994) using an H~I 21~cm emission line map.

We have examined variations in shifts, widths and intensities of the
stronger narrow lines between different species and epochs.  The lines
considered were H$\alpha$,$\beta$,$\gamma$, He~I 4471, 5015, 5875~\AA,
[O~III] 4959, 5007~\AA, [NII] 5755~\AA, [Ne~III] 3869~\AA\ and
[Fe~III] 4658, 5270~\AA.  In Figure~\ref{fighivel} we show in detail
those lines which were detected on both days~17.4 and 36.3 {\it viz.}
H, He~I and [O~III].  The relative velocity shifts discussed below
are with respect to our adopted SN centre of mass velocity of
+847~km/s.

The following general conclusions can be reached.  Where the S/N was
sufficiently high, we can say that the peaks and troughs of the H and
He lines exhibited similar velocity shifts with respect to the SN
centre of mass frame.  On day~17.4, their emission peaks were at
+8$\pm$3~km/s, moving to +15$\pm$3.5~km/s on day~36.3.  The red wings
of the H lines appear to shift to the red by around 15~km/s between
the 2 epochs. Similar behaviour may be apparent in some of the He~I
lines also.  In contrast, the absorption troughs (which also showed no
difference in velocity shift between H and He) remained at a mean
shift of --25$\pm$3~km/s between the two epochs.  The blue extremities
of the H lines extended to $\sim$--55~km/s on both epochs, whereas the
blue limit of the He~I lines was $\sim$--40~km/s.  The red extremities
of the H and He~I lines lay at $\sim$+60~km/s apart from H$\alpha$
which reached about +90~km/s.  Figure~\ref{fighivel} also reveals that
the absorption component in most of the H and He~I lines became
significantly deeper, while the emission component showed a much
smaller intensity variation.  Only in H$\alpha$ did the depth of the
absorption remain about the same.  (This may also apply to He~I
4471~\AA\ but the S/N is very low.)  On day~17.4, the only detectable
forbidden lines were those of [O~III] and only [O~III]~5007~\AA\ was
of good S/N.  The blue edge of this line lay at $\sim$--40~km/s,
similar to that of the He~I lines.  However the red edge extended no
further than +15~km/s.  By day~36.3, the blue edge remained at
--40~km/s, but the red edge had moved to $\sim$+40~km/s.  As mentioned
above, we suggest that this is due to light travel time effects.  By
this epoch, lines of [NII], [Ne~III] and [Fe~III] were also measurable
and these all showed about the same absolute redshift and width as the
[O~III] lines.  The [O~III] profiles showed little variation in peak
flux between the two epochs.

In summary, the red wings of most of the emission lines showed a
movement to the red of around +10--+20~km/s between the two epochs,
while the blue limits of the forbidden line emission profiles and the
allowed line absorption profiles remained roughly stationary i.e. the
extent of the lines effectively increased by +10--+20~km/s between the
two epochs due the redshift of the red wing.  We attribute this to
light travel time effects.  We also note that the fact that the
[O~III]~5007~\AA\ profile varies in time, together with the similarity
of its shape and redshift on day 36.6 to those of other forbidden
lines, indicates that the SN CSM is indeed the origin of these
lines. This rules out the alternative line-of-sight emission line
region suggested by Bowen {\it et al.} (2000).  

The narrow P~Cygni lines of He~I 10830~\AA\ acquired at UKIRT
(resolution $\sim$100 km/s) on days~44 and 55 are also listed in
Table~\ref{fighires_id}.  The emission profiles and peak-trough
profiles are unresolved.   The emission peak lies at velocities of
+29$\pm$20 km/s on day~44 and $-7\pm$15~km/s on day~55, which is the
same as for the optical He~I lines, to within the uncertainties.  The
line strengths are virtually identical at the two epochs.  While the
main emission and absorption components are unresolved, there is some
indication that the blue wing of the absorption is barely resolved,
reaching a blueshift of $\sim$--300~km/s {\it i.e.}  similar to that
found for the wing of the broad H$\alpha$ absorption.

We conclude that the narrow lines arose in a CSM wind of velocity of
40--60 km/s.  Highest velocities ($\sim$60~km/s) are seen in the H
lines, and lowest ($\sim$40~km/s) in the forbidden lines.  The greater
velocity extent of the H lines may be in part due to the stronger
effects of thermal broadening in this element.  For a temperature of
10,000~K, the thermal broadening contribution has a FWHM of 21~km/s in
hydrogen compared with just 5~km/s in oxygen.  It is also possible
that radiative acceleration effects could contribute to the greater
width of the allowed lines.  Acceleration of the CSM could be driven
by the UV radiative precursor and/or cosmic ray precursor (Fransson,
Lundqvist \& Chevalier 1996), with the innermost part of the CSM being
accelerated most.  The optical depth in allowed lines in the
accelerated CSM component could be large (due to high radiative
excitation) thereby increasing the width of the scattered lines.
However, it is quite plausible that the volume emission measure of the
accelerated component is low and so would not contribute to the
forbidden lines.  

Turning briefly to the broader absorption component in the optical H
and He~I lines, we can see (Figure~\ref{fighivel}) that this feature
strengthened somewhat between days~17.4 and 36.3.  Similar
behaviour is reported for the H lines by Bowen {\it et al.} (2000).
However, Bowen {\it et al.} also note the presence of a similarly
broad absorption in Mg~II 2796, 2803~\AA\ which (a) may have {\it
declined} in strength between days 32 and 39, and (b) never exhibited
narrow P~Cygni lines.  We also note that the H$\alpha$ and H$\beta$
lines are sufficiently well defined to allow us to conclude that the
broad minima were blueshifted by $\sim$--115~km/s on day~17.4, moving
redward to $\sim$--85~km/s by day~36.3.

\section{Discussion}
\subsection{The circumstellar medium of SN 1998S}
The CSM of SN~1998S has been described and discussed by a number of
authors (L00, G00, Fassia {\it et al.} 2000, Bowen {\it et al.}  2000,
Lentz {\it et al.} 2000).  In the following discussion we adopt the
general consensus about the form of the CSM and the evolution of the
spectra.  In particular, we confirm the two-phase CSM proposed by L00,
but we deduce somewhat different, and occasionally firmer constraints
on the size and density of the CSM.  In this paper we focus mostly on
what can be deduced from the first-season observations.  Later
observations will be described in a subsequent paper (Fassia {\it et
al.}, in preparation).  We envisage that before it exploded, the
progenitor star of SN~1998S had undergone two phases of mass loss,
with a gap or a phase of reduced mass loss in between.  This gave rise to two
shells of CSM.  Following L00, we shall refer to these as the Inner
CSM (ICSM) and Outer CSM (OCSM).  The interaction of the ejecta with
these two shells is responsible for the early ($t<$10days) and late
($t>$40days) time appearance of broad emission lines lacking a P Cygni
absorption component ({\it cf.} Chugai \& Danziger 1994, Chevalier \&
Fransson 1994).

\subsubsection{The inner circumstellar material (ICSM).}
We recall that the early-time broad component had a
FWHM$\geq$4000~km/s, was present as early as day~3, and had faded
significantly by day 12--14 (Qiu {\it et al.} (1998); this work).
While lines of this width in SNe are usually associated with the
ejecta, the absence of any corresponding P~Cygni absorption is a
robust indicator that these lines are driven by the high energy
radiation of the shock resulting from the interaction of the ejecta
with dense CSM in the immediate vicinity of the supernova (Chugai
1990).  The ratio of the luminosities of the H$\alpha$ and H$\beta$
broad components in the first week spectra is
$f_{H\alpha}/f_{H\beta}$=1.45$\pm$0.25 and the ratio of the fluxes of
the P$\beta$ and H$\gamma$ lines is
$f_{P\beta}/f_{H\gamma}$=0.22$\pm$0.07. If we take into account the
reddening assuming $A_{V}=0.68^{+0.34}_{-0.25}$ (Fassia {\it et al.}
2000) then these ratio become $f_{H\alpha}/f_{H\beta}$=1.1$\pm$0.7,
$f_{P\beta}/f_{H\gamma}$=0.1$\pm$0.06. These low ratios indicate a
high degree of thermalisation for these lines and thus imply a high
density and a large optical depth for the line-forming region.

The fading of the broad emission components by days~12--14.3 and the
appearance of broad absorption components in H and He~I by days
14.3--17.7 ({\it cf.} Figure~\ref{opspec_ncont}) is strong evidence
that by that time the ejecta had completely overrun the ICSM.  The
highest ejecta velocity was about $\sim$10,000~km/s relative to the SN
center of mass. We therefore estimate that the outer limit of the ICSM
was $\leq$90~AU.  By $\sim$15~days the photospheric radius was already
$\sim$75~AU (Fassia {\it et al.} 2000).

\subsubsection{The region between the ICSM and OCSM.}
As can be seen in Figures~\ref{opspec_ncont}, \ref{irspec_ncont} the
spectra during 14--27~days are dominated by broad P~Cygni-type
absorption profiles. During this phase, the extreme blue edge of the
broad absorption indicated a velocity of --8,000~km/s relative to the
SN centre of mass.  We adopt this as the maximum velocity of the
ejecta.  The ejecta collided with the OCSM sometime after day~27.  The
high resolution spectra of day~36 show no sign of broad emission.
However, by day~44, strong broad emission had appeared in the IR
lines.  We deduce that the 8,000~km/s shock reached the inner edge of
the OCSM on day 40$\pm$4 indicating a radius of 185$\pm$20~AU.  So the
region between the ICSM and OCSM extended from $\leq$90~AU to
$\sim$185~AU.

There is evidence that even during the phase when the ejecta had
overrun the ICSM but not yet reached the OCSM, the ejecta/CSM
interaction contributed significantly to the spectrum.  While the
SN~1998S spectral lines at this phase are similar to those observed in
the spectra of normal SNe~II during their photospheric phase
(eg. Blanton {\it et al.} 1995, Fassia {it et al.} 1998), the SN~1998S
spectrum differs in that it has a much stronger, bluer continuum and
shallower absorption profiles.  As mentioned in Section~3.1.3, Branch
{\it et al.} (2000), Lentz {\it et al.} (2000) and L00 suggest that
the low contrast of the broad lines relative to the continuum is due
to external illumination of the line-forming region by light from the
CSM-ejecta interaction. We agree with this suggestion and furthermore
propose that the observed strong continuum emission arises in a
massive, cool, dense shell of ejecta formed at the ICSM-ejecta
discontinuity (e.g. Chevalier \& Fransson 1994).  Such a shell is
likely to be produced by the interaction of the ejecta with the ICSM,
and would have a velocity jump of $\sim10^3$ km s$^{-1}$ between the
inner part of the ejecta and the shell.  This could have led to
enhanced continuum emission in two ways. In the first mechanism, even
after the ejecta have completely overrun the ICSM, the existence of
the velocity jump would have resulted in a reverse shock wave which
ran through the ejecta and which would have remained powerful enough
to generate strong continuum emission from the dense shell.  For the
second mechanism we suppose that between the ICSM and the OCSM there
actually existed a tenuous, uniformly distributed circumstellar medium
which we shall to refer to as the MCSM (Mid-CSM).  Interaction of the
dense, swept-up shell with the MCSM would have released power (kinetic
luminosity) according to $ L= 2\pi R^2 \rho v^3$ or $L\propto R^2
\propto t^2$.  Interestingly, between days 17.7 and 44.6 the P$\beta$
flux grew $\sim$6 times, in good agreement with the expansion factor
between these two epochs {\it viz.}  $(t_2/t_1)^2=6.3$.  We note,
however, that if the P$\beta$ lines arose from the dense shell and
were thermalized (saturated) at a constant excitation temperature then
the expansion of the dense shell alone could account for the $L\propto
R^2$ behaviour.  Nevertheless, if the dense shell did exist then it
may have contributed to the broad line emission which appeared in the
IR by day~44 {\it i.e.} at this epoch the line luminosities may have
been composed of contributions from both radioactive and ejecta/CSM
energy sources, where ``CSM'' includes ICSM, MCSM and OCSM.  However,
determination of the relative luminosities from these sources, and
what fraction of the ejecta/CSM luminosity arose from the different
zones, is beyond the scope of this paper.

\subsubsection{The outer circumstellar material (OCSM).}
The presence of the OCSM is revealed in two phases.  The earlier is
due to its interaction with the UV/X-ray flash from the SN, and the
later is due to its interaction with the SN ejecta. \\

{\it a) The early phase.} \\ We attribute the early, narrow, usually
unresolved, lines to recombination and heating following ionisation of
the slowly-moving OCSM by the UV/X-ray flash at shock break-out.
These lines were present in the earliest spectra (day~3.3) as the
narrow (FWHM$\leq$500~km/s) components of the emission lines.  By
days~12--14, the narrow H, He~I lines began to take on the form of
P~Cygni profiles, and this is clearly confirmed in the day~17.4 WHT/UES
spectra.  However, the high resolution spectra also show that the
narrow lines in H and He~I are actually made up of at least two
components.  The components are best described as a P~Cygni-like
profile of maximum velocity $\sim$50--60~km/s and a broader,
blueshifted absorption extending to around --350~km/s.  The high
resolution spectra also show narrow forbidden lines extending to
$\pm$40~km/s of the centre of mass velocity.

We consider first the narrowest lines or line components i.e. those
extending to $\pm$40--60~km/s.  We envisage that the narrow
P~Cygni-like profiles in H and He~I were produced by two effects.  The
recombination cascade in the OCSM would generate narrow emission
lines.  In addition, the recombination would populate excited levels
in the OCSM, resulting in photospheric photons being resonantly
scattered, hence forming classical P~Cygni lines.  Thus, the narrow
line emission component comprises photons emitted both through
recombination and resonant scattering.  We note that the He~I 6678,
7065~\AA\ lines appear only in emission and only in the 17.4 day
spectra.  He~I 6678~\AA\ is produced by a singlet transition to the
$2p^1P_0$ level. The lack of an absorption in this line is presumably
due to rapid decay from $2p^1P_0$ to the metastable $2s^1S$ level. In
contrast, the 5015~\AA\ line is produced by a transition directly to
the $2s^1S$ level, and that accounts for the absorption in this line.
Thus, He~I 6678~\AA\ is a virtually pure recombination line and its
disappearance by day~36.3 is presumably due to a lowering of the
recombination rate by then.  He~I 7065~\AA\ is produced by a triplet
transition to the $2p^3P_0$ level.  The presence of absorptions in
the 5876 and 4471~\AA\ lines (which both also decay to the $2p^3P_0$
level) indicates a significant population in $2p^3P_0$.  However, the
cross-section for resonant scattering from this level via the
7065~\AA\ transition is only about 10\% and 50\% relative to that of
the 5876~\AA\ and 4471~\AA\ transitions respectively. Given that the
4471~\AA\ line barely exhibits an absorption component above the
noise, it is not surprising that there is little evidence of
absorption in the 7065~\AA\ line.  The stronger pronounced emission
component in 7065~\AA\ relative to 4471~\AA\ may be due to its being
of lower excitation energy.  Collisional excitation from the
metastable $2s^3S$ level would thus be expected to produce a
significantly stronger 7065~\AA\ line.  The disappearance of the
7065~\AA\ line by day~36.3 is probably due to cooling of the electron
gas.The forbidden lines would have been excited as a result of the
heating of the OCSM electron gas following recombination.  The
appearance of narrow [N~II], [Ne~III] and [Fe~III] lines between
days~17.4 and 36.3 can be attributed to the formation of the relevant
species following recombination from higher ionisation.

We can estimate the CSM density in the region where the [O~III] lines
were formed by using the ratio of their fluxes
R=$(f_{\lambda4959}+f_{\lambda5007})/f_{\lambda4363}$. L00
measured the ratio of the narrow [O~III] lines present in their 4d and
5d spectra days and found that for temperatures in the range
15000~K$<$T $<$ 30000~K the electron density of the CSM was $6\times
10^{6}<n_{\rm e}<2\times 10^{7}$cm$^{-3}$.  However, the SPSDAS
package of IRAF that they use to derive these values is appropriate
only for a low density (n$\lesssim 10^{5} cm^{-3}$) plasma. In
addition, theoretical models developed to study the temperature and
structure of the CSM around supernovae (Lundqvist \& Fransson 1988)
have shown that the CSM temperature is very sensitive to the ionising
spectrum of the supernova and varies both with radius and in time. The
temperature range considered by L00 for the O III-rich gas is much
narrower than in the models of Lundqvist \& Fransson (1988), which
indicate 10$^{4}$--10$^{5}$~K for a distance greater than $\sim$200~AU
in a situation similar to that of SN 1998S. The upper limit,
$10^{5}$K, is set by the fact that O~III is collisionally ionised to
O~IV at higher temperatures.

To calculate the CSM density from the [O~III] line ratio we therefore
assumed that $10^{4}~K<T < 10^{5}~K$ and used the 6-level [O~III]
model atom described in Maran {\it et al.} (2000).  This model
includes O~III atomic data from Aggarwal (1993) and Galavis {\it et
al.} (1997) and takes into account collisional de-excitation that
plays an important role for $n\gtrsim10^{5} cm^{-3}$.  Unfortunately,
our 3.3~d--17.7~d low-resolution spectra are of insufficient S/N for
us to estimate the ratio, R, of the oxygen lines. However, we were
able to estimate the ratio on day~36.3 using our high resolution
observations.  Before de-reddening, the ratio is R=$3.3\pm1.0$.
De-reddening using the Cardelli {\it et al.}  (1989) law with
$A_{V}=0.68^{+0.34}_{-0.25}$ (Fassia {\it et al.} 2000), changes the
ratio to R=2.9$\pm$1.5. Therefore, using the 6-level model atom
mentioned above we estimate that for 1.4$<$R$<$4.4 the lower limit to
the CSM density is $1\times 10^{6}$ cm$^{-3}$.  A CSM of this density
and velocity (40--50~km/s) is consistent with it being a red
supergiant wind (Dupree 1986, Salasnich, Bressan \& Chiosi 1999, Jura
\& Kleinmann 1990).  No solution for the density is found for $T \leq
8000$~K (see Fig. 12), which means that this is a firm lower limit to
the temperature in the O III-rich gas at this epoch. If the
temperature is $T \leq 11000$~K, which cannot be excluded from the
results of Lundqvist \& Fransson (1988), no upper limit to the density
can be given. The day~36.3 line ratio therefore only places lower
limits on the density and temperature of the O III-rich gas.

\begin{figure}
\vspace{8.2cm}
\includegraphics{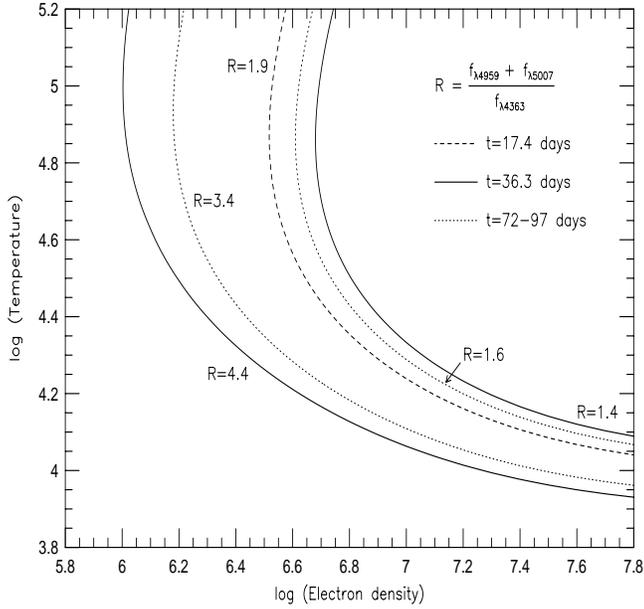}
\caption[]{Solutions of the 6-level model atom for  values of the [O~III]
line ratio, R, estimated from the SN 1998S spectra (cf Section 4.1.3)}
\label{figoxyg}
\end{figure}

For 17.4~d, only an upper limit is available for the flux of 
[O~III]~4363~\AA\ from our high resolution spectra and so we get R$>$
1.9 (dereddened).  From Figure~\ref{figoxyg} we see that this, unfortunately,
does not constrain the density for the temperature range
10$^{4}$-10$^{5}$~K.

We turn now to the broader ($\sim$350~km/s) shallower absorption
component seen in the H and He~I lines.  We recall that this feature
deepened between days~17.4 and 36.3.  The minima also moved redward
from $\sim$--115~km/s to $\sim$--85~km/s.  The discrete difference in
its FWHM, relative to that of the narrow component might indicate that
the two components were formed in different regions.  The presence of
the broader absorption as late as day~35 implies that it too must have
arisen in the OCSM.  It could not have been due to the ICSM since it
had been totally overrun by then.  In addition, an origin in the
region between the ICSM and OCSM seems unlikely since not only must
the density of material there have been quite low (see above) but also
by day~35 the undisturbed region between the shock and the inner edge
of the OCSM was very thin - barely $~\sim$25~AU.  While a wind
velocity $\sim$350~km/s is too high for a red giant, blue supergiant
(BSG) winds do exhibit velocities in the range $\sim$100~km/s to
$\sim$3000~km/s (eg. Abbott {\it et al} 1981).  Thus a possible
explanation for the broader absorption is that it occurred in a
component of the OCSM produced when the progenitor was going through a
BSG phase. The deepening and shift to the red suggests the effect of a
BSG wind ``piling up'' against the inner boundary of the RSG wind.
This would produce a negative velocity gradient in the BSG wind, as
well as making its density gradient less steep.  As the SN shock
progressed, it would first engulf the faster, more tenuous part of the
BSG wind, leaving the outer, slower-moving zone undisturbed, thus
accounting for the decrease in the absorption blueshift. The deepening
absorption could be due to an increase in the density of the outer,
slower-moving zone of the BSG wind as it became increasingly
compressed between the shock and the RSG wind boundary.  The division
of the OCSM into two zones in this scenario would also provide a
natural explanation for the apparently discrete difference in the FWHM
of the narrower and broader absorption features.

A possible alternative explanation for the $\sim$350~km/s absorption
component is that it might be produced by the acceleration of the
innermost OCSM by the radiation pressure of the UV precursor,
mentioned at the end of Section~3.3.  This would also produce the
required negative velocity gradient.  Details of the physics of the
high velocity wind component will be considered in a separate paper.

{\it b) The later phase} \\ As indicated above, we take the presence
of strong, broad Paschen and Brackett lines on day~44 to imply that
the shock/OCSM interaction had already already begun by this date.
This is further supported by the weakening of the narrow components of
the H and He~I lines (see Figs. 9 and 10), which we attribute to the
invasion of the OCSM by the shock.  The broad lines persisted to at
least day~692 (Fassia {\it et al.} in preparation) with material still
moving as fast as 4600~km/s relative to the SN centre of mass at that
time.  From this we infer that the OCSM extended at least as far as
1800~AU from the supernova. For a wind velocity of 50~km/s, this
implies that the the ejection of the OCSM began at least 170 years
ago, in agreement with L00.  We agree with L00 that line ratios at
this later time continue to imply a high density in the line formation
region of the OCSM.  Our estimate for the deredened [O~III] ratio is
$(f_{\lambda4959}+f_{\lambda5007})/f_{\lambda4363}=2.5\pm0.9$ (average
of values from days 72 and 97 ) which suggests (see
Figure~\ref{figoxyg}) densities $ n_{e} > 1.5\times 10^{6}cm^{-3}$ for
a temperature of $10^{4}~K<T <10^{5}~K$, consistent with the density
derived at earlier epochs by us and by L00.

From the above we can obtain lower limits for the OCSM mass and for
the mass loss rate that produced it.  Assuming that the OCSM has the
form of a spherically symmetric wind of constant velocity $v_w$.  the
density would vary as $\propto r^{-2}$.  Consequently the mass of the
OCSM is:
\begin{equation}
M\gtrsim 4 \pi r_{in}^{3} m_H n \mu (\frac{r_{out}}{r_{in}}-1)
\end{equation} 
where $\mu$ is the mean atomic weight in amu, $m_{H}$ is the mass of
the hydrogen atom, $r_{out}, r_{in}$ are the inner and outer limits of
the OCSM and $n$ is the total number density at the inner limit.
Making the conservative assumption of complete ionisation and that the
OSCM consists mostly of hydrogen, we have $n \approx 2n_{e}$ and
$\mu$=0.5.  Therefore, substituting $n_{e}\sim1.5\times10^{6}$
cm$^{-3}$, $r_{in}\sim185$ AU and $r_{out}\gtrsim1800$ AU, we have
$M\gtrsim0.003 M_{\odot}$. For a wind velocity of 50~km/s and a
constant mass loss we obtain a mass loss rate of $\gtrsim 2 \times
10^{-5} M_{\odot}/yr$ taking place for at least 170~years before the
explosion.

\subsection{Carbon Monoxide}
Molecular emission from a supernova was first observed in the form of
CO emission from SN~1987A (Catchpole \& Glass 1987; McGregor \& Hyland
1987).  Measurement of the first overtone spectrum as early as day~110
was reported by Meikle {\it et al.} (1989).  However, it is possible
that CO formed even earlier.  The sudden rise in the $K-M$ colour at
about day~100 (Bouchet \& Danziger 1993) may have been due to CO
fundamental emission in the $M$-band.  The evolution of the SN~1987A
first overtone emission spectrum was monitored by Meikle {\it et al.}
(1989, 1993) to 1.5~years.  Other species responsible for molecular
emission from SN 1987A include SiO together with more tentative
identifications of CS, FeO, H$_3^+$ and HeH$^+$ (Dalgarno, Stancil \&
Lepp, 1997, and references therein).  Of these molecules, emission
from CO has been most extensively studied.

Analysis of the first overtone CO band in SN~1987A has demonstrated
the value of this type of study in determining the velocity,
excitation temperature and CO mass in the CO-emitting zone.  The first
models (Spyromilio {\it et al.} 1988), which assumed LTE, provided
good fits for days~192 \& 255 implying a CO mass of
$\sim$10$^{-4}$~M$_{\odot}$ Velocities of 2000~km/s and 1800 km/s,
respectively, were inferred for the two epochs.  At later epochs
poorer fits were obtained. This was attributed to deviation from a
Boltzmann distribution.  Subsequent interpretations by several groups
involved non-LTE modelling for the vibrational transitions. An example
is the work of Liu \& Dalgarno (1995) who obtained much-improved fits
for the late-time data.  They inferred a mass of CO as high as
0.04~M$_{\odot}$ on day~110 but which was reduced quickly to
$\sim$3$\times$10$^{-3}$~M$_{\odot}$ by day~250 due to destruction by
the fast electrons. They found that, when present, the vibrational
transitions of CO provide the dominant cooling mechanism of the C/O
zone during the first 2~years, reducing the temperature to $\sim$700~K
by the end of that period.  Prior to SN~1998S, the only other
supernova (besides SN~1987A) for which CO emission has been detected
is the type~II SN~1995ad at around 105~days, observed by Spyromilio \&
Leibundgut 1996.  Following Liu \& Dalgarno, they suggest that a CO
mass as high as 0.05~M$_\odot$ might have formed.  In SN~1998S intense
first overtone CO emission spectra were measured at two epochs (days
110 and 240, with our definition of t$_{0}$= 1998 March 2.68 UT) by
G00 and at a single epoch (day 109) by us.  We shall compare their
conclusions with ours at a later point in this section.

\subsubsection{Non-LTE model of CO emission.}
To interpret the CO spectrum of SN~1998S, we have constructed a model
following a similar non-LTE approach to that of (Liu \& Dalgarno
1995), including CO formation and gamma-ray energy deposition.  The
only additional feature is a simple treatment of non-local radiative
coupling of the first overtone lines in a clumpy envelope.  By
adjusting the model to reproduce our day~109 CO spectrum
(Fig.~\ref{irspec_ncont}, \ref{co}) we
can estimate the mass, density, temperature and velocity of the CO and
C/O-rich matter as well as the mass of $^{56}$Ni.  To compare the
model with the data, the model luminosity was converted to flux
assuming a distance of 17~Mpc (Tully, 1998).

In the model we assume that the CO emission arises from an ensemble of
C/O-rich clumps embedded in a homogeneous (on average),
macroscopically-mixed core.  All the C/O-rich clumps are composed of
equal proportions by mass of C and O, and have similar size and
physical conditions.  Following Liu \& Dalgarno (1995) we assume that
the total mass of the clumped C/O-rich mixture is $M_{\rm cl}=0.2
M_{\odot}$.  This is reasonably consistent with mass estimates from
stellar evolution computations for stars in the range 12--25
$M_{\odot}$ (Heger 1999).  Gamma-ray energy deposition is determined
using the escape probability approximation with an absorption
coefficient of $k_{\gamma}=0.03$ cm$^2$ g$^{-1}$.  However, it is
likely that the $^{56}$Ni is also clumped. Consequently, as the
gamma-rays travel from the $^{56}$Ni clumps to the C/O-rich clumps
they will suffer attenuation, the amount of which will decrease with
age.  This process has been invoked in SN~1987A to account for the
slow evolution of the CO luminosity at early epochs (Liu \& Dalgarno
1995).  To allow for this effect in SN~1998S, we follow Liu \&
Dalgarno by including an energy reduction factor $\psi$ by which the
deposition rate in the C/O-rich material is multiplied.  To estimate
$\psi$ for SN~1998S on day~109, we compared the first overtone CO flux
with that reported by G00 on day 225.  The density of the C/O-rich
clumps ($\rho_{\rm cl}$) may be higher than the average density in the
core ($\rho_{\rm c}$), and so we specify the overdensity with a
density contrast parameter $\chi=\rho_{\rm cl}/\rho_{\rm c}\geq 1$.
The CO concentration is primarily controlled by CO formation via the
radiative association $\rm{C}+\rm{O}\rightarrow\rm{CO}+h\nu$ and
destruction via ionization by fast electrons
$\rm{CO}+e\rightarrow\rm{CO}^{+}+e$ followed by the dissociative
recombination $\rm{CO}^{+}+e\rightarrow\rm{C}+\rm{O}$ (Liu \& Dalgarno
1995).

The fraction of deposited energy channelled into Coulomb heating was
obtained using calculations of Kozma \& Fransson (1992).  The electron
temperature ($T_{\rm e}$) of the C/O-rich clumps was found assuming
energy balance between the energy deposited as heat by Compton
electrons and the radiation emitted in the CO fundamental and first
overtone bands.  Cooling in [OI] 6300, 6364 \AA\ lines was found to
contribute typically less than $4\%$ of that due to the CO bands, in
agreement with the analysis by Liu \& Dalgarno (1995).

The non-LTE population of vibrational levels was calculated assuming
that radiative and collisional transitions between neighbouring
vibrational levels are dominant.  This is a sensible approximation
since the radiative transition rates for the first overtone are an
order of magnitude smaller than those for the fundamental.  The
electron collision rate coefficients were taken from Liu et
al. (1992), while oscillator strengths are from Kirby-Docken \& Liu
(1978).  Twelve vibrational states, with 101 rotational levels for
each vibrational state, were considered.  This yields a line list of
2000 lines for the first overtone band and 2200 lines for the
fundamental band.  The CO radiative rates are affected by continuum
radiation originating outside the clumps.  The effects of this were
determined by assuming that the core luminosity is equal to the
overall deposited radioactive luminosity.  The external continuum
spectrum was approximated by a dilute blackbody with a colour
temperature of 5000~K.  In practice, the model is not very sensitive
to the choice of the temperature since the effect of the external
continuum radiation is small compared to that of the collisions.  This
is in accordance with the Liu \& Dalgarno (1995) conclusions about
SN~1987A.  After the populations of the vibrational levels were found,
the rotational levels of individual vibrational states were populated
according to the Boltzmann distribution.

Radiative transfer in the clumpy CO matter was treated in terms of the
average escape probability.  This includes the combined effects of the
escape probability of photons from an average CO cloud, together with
absorption by the other CO clouds in the core.  Our models reveal that
variation of the CO cloud radius between 10~km s$^{-1}$ and 100~km
s$^{-1}$ (assuming homologous expansion) only slightly affects the
shape of the spectrum.  We adopted a cloud radius of $a=50$~km
s$^{-1}$, very similar to the oxygen clump size estimates for SN~1987A
(Stathakis {\it et al.} 1991).  The weak sensitivity of the spectrum shape
to the cloud size is due to the opposing trends of photon absorption
in the CO emission source cloud and in other clouds.

The final model is specified using four free or ``input'' parameters:
core velocity ($v_{\rm c}$), core mass ($M_{\rm c}$), density contrast
($\chi$), and $^{56}$Ni mass ($M_{\rm Ni}$).  It should be stressed
that not all these parameters are equally constrained by the
observations.  In particular, good fits over a wide core mass range
may be achieved by simply adjusting the other three parameters
appropriately.  However, we find that models with a core mass $M_{\rm
c}<0.4~M_{\odot}$ require an implausibly high density contrast in the
C/O-rich matter ($\chi \geq 10$).  On the other hand, models with a
core mass $M_{\rm c}>4~M_{\odot}$ require a $^{56}$Ni mass
$>0.2~M_{\odot}$, inconsistent with the bolometric luminosity of
SN~1998S around day 130 (Fassia {\it et al.} 2000).  We have therefore
considered the two extreme cases set by these considerations {\it
viz.} model~A with a core mass of $0.4~M_{\odot}$ and model~B with a
core mass of $4~M_{\odot}$.

\subsubsection{Comparison of the CO model with the day~109 spectrum.}
\begin{figure}
\vspace{9.5cm} \includegraphics{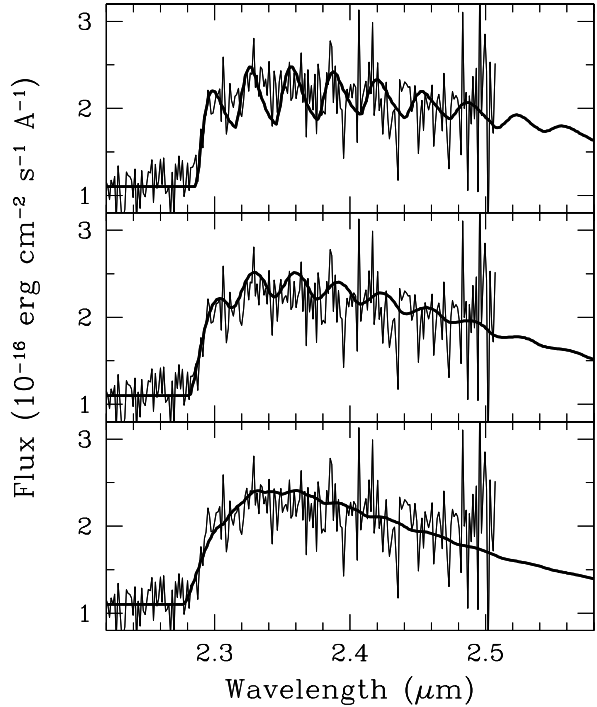} \ 
\caption[]{ Carbon monoxide first overtone model~A (thick line) ({\it
cf.} Table~\ref{table_co}) compared with observations (thin line).  Models are shown
for three CO emission zone velocities: 1600 km s$^{-1}$ (upper panel),
2200 km s$^{-1}$ (middle) and 2800 km s$^{-1}$ (bottom).}
\label{co}
\end{figure}

We find that both models are able to reproduce the observed CO first
overtone emission for plausible combinations of the free parameters
$v_{\rm c}$, $\chi$ and $M_{\rm Ni}$.  From these models we obtain
physically reasonable values for the ``output'' parameters $M_{\rm
CO}$, $T_e$, $n_b$ $(n_{\rm b}$ is the density of the C/O-rich matter
expressed as a baryon number density).  In Fig~\ref{co} we show
model~A fits for core velocities of 1600, 2200 and 2800~km/s.  We can
immediately eliminate the two extreme values.  In the 1600 km/s model,
the visibility of the quasi-periodic structure produced by the
R-branches of the 0--2 and 1--2 vibrational transitions is too high.
In the 2800 km/s model the visibility is too low and the rising blue
edge too shallow.  We deduce that the velocity of the CO emission zone
is $2200\pm300$ km s$^{-1}$.  A similar result is obtained for
model~B.  For the optimum velocity of 2200~km/s, in
Table~\ref{table_co} we show the input and output parameters for the
two models.

We find that $n_{\rm b}$, $M_{\rm CO}$ and $T_e$ are quite
similar for models~A and B.  To narrow down further the range of
models encompassed by A and B, additional information is required.
One such fact is the estimate by Fassia {\it et al.} (2000) of a
$^{56}$Ni mass of $0.15\pm0.05~M_{\odot}$, based on the bolometric
light curve of SN~1998S.  This is consistent with model~B but not
model~A (Table~\ref{table_co}).  A caveat is the possible contribution to the
luminosity from the ejecta-wind interaction which might decrease the
$^{56}$Ni mass implied by the bolometric light curve.  However, Fassia
{\it et al.} point out that the observed bolometric decline rate matches
well the radioactive luminosity decay rate, arguing against a
significant ejecta-wind contribution to the luminosity.  Another
argument in favour of model~B is based on the issue of the energy
reduction factor.  The model~A fit requires $\psi=0.2-0.3$,
indicating a substantial difference between the average intensity of
gamma-rays in the $^{56}$Ni zone and C/O-rich clumps.  Yet, in this
model the optical depth of the core for gamma-rays is only about 1.
Moreover, macroscopic mixing should lead to even lower optical depths
in the 'walls' between $^{56}$Ni and C/O zones.  These low optical
depths are inconsistent with a $\psi$ value as low as 0.2--0.3, as is
required to fit the data.  On the other hand, the version of model~B
rm which fits the observed spectrum has a core optical depth of $\sim$10
on day 109, consistent with its $\psi$ value of $\sim0.5$ (Table~\ref{table_co}).
\begin{table*}
\caption{Carbon monoxide model parameters}
\medskip
\begin{tabular}{ccccc|ccc}
\\
& \multicolumn{4}{c|}{Input parameters (see text)} & \multicolumn{3}{c}{Output
parameters (see text)} \\ 
\\
Model & $M_{\rm c}$ & $M_{\rm Ni}$ & $\chi$ & $\psi$ & $n_{\rm
b}$ & $M_{\rm CO}$ & $T_{\rm e}$ \\
      & ($M_{\odot}$) & ($M_{\odot}$) & & & ($10^{11}$ cm$^{-3}$) &
($10^{-3} M_{\odot}$) & (K) \\ 
\\
A     &   0.4   &  0.07   &  8  &  0.24  &  0.77  &  1.2  &  3700  \\
B     &    4    &  0.2    &  1  &  0.52  &  0.97  &  1.3  &  3650  \\
\\
\end{tabular}
\label{table_co}
\end{table*}

In summary, we favour the high mass core model (model~B) with a core
mass of $4~M_{\odot}$, core velocity of 2200$\pm$300 km/s, electron
temperature of 3650$\pm$200~K, negligible overdensity of the C/O-rich
clumps (density contrast parameter $\chi\sim$1) and a $^{56}$Ni mass
of about 0.2~M$_\odot$.  (Note that the C/O is still clumped in the
sense of chemical abundance.)  Unfortunately, the mass of the mixed
core does not permit us to draw definite conclusions about the
progenitor mass.  The reason is that we lack information about the
relative proportions of metal-rich and H/He-rich materials in the
core.  That the amount of H/He-rich material might be significant is
supported by the observation of sharp-topped H$\alpha$ and He
I~10830~A line profiles at $t\approx 100$ days, indicating the
presence of low velocity hydrogen and helium ($v\approx 1000$ km
s$^{-1}$).

\begin{figure}
\vspace{9.5cm} \includegraphics{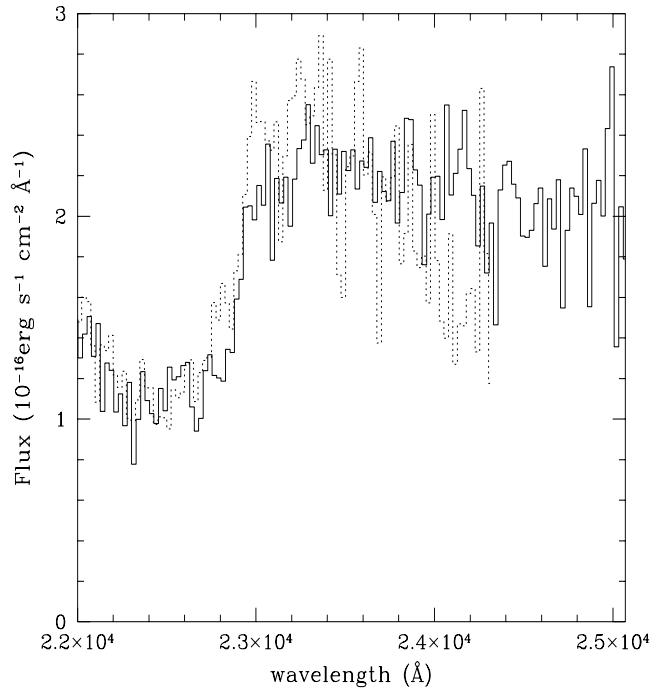} \ 
\caption[]{Comparison of the first overtone CO spectrum obtained on
day~109 (solid line) with the spectrum obtained by G00 on day~110
(dotted line).  The spectra have not been corrected for the redshift
of the supernova.}
\label{co_ger}
\end{figure}

G00 have analysed the CO emission from SN~1998S covering about the
same epoch as did our study.  Matching their day~110 CO first overtone
spectrum with a range of models they deduce a CO velocity of
2000--3000 km/s.  They argue that a CO velocity of 3000 km/s would
imply a progenitor mass in excess of 25~$M_\odot$, which is outside
the plausible progenitor mass range for our adopted C/O-rich core
mass.  However, we believe that our analysis eliminates the upper half
of their velocity range, which presumably would reduce the inferred
progenitor mass.  But in any event, as pointed out above, there are
difficulties in reaching firm conclusions about the progenitor mass.
We note that G00's CO velocity value depends quite strongly on the
presence in their spectrum of a ``smooth rise to the blue edge of the
CO feature''.  However, our spectrum, taken within a day 
of their's does not show this smooth rise.  Instead, our spectrum
remains roughly flat to about 2.27~$\mu$m and then rises abruptly.  In
Figure~\ref{co_ger} we compare the two spectra, binned to 25~\AA\ per
pixel.  It can be seen that the agreement is excellent up to
2.27~$\mu$m, but then the spectra become more discrepant.  In
particular, the G00 spectrum exhibits a ``bump'' at about 2.28~$\mu$m
which contributes significantly to the perception of their smooth rise
to the blue edge. However, their spectrum appears to be rather noisy
redward of 2.3~$\mu$m and it may be that noise is also responsible for
their 2.28~$\mu$m bump.  Our value for the electron temperature,
3450--3850~K, also confines the 3000--4000~K range suggested by G00.

As indicated above, Liu \& Dalgarno (1995) deduce a CO mass of $\sim
10^{-2}~M_{\odot}$ in SN~1987A at 110 days, which is an order of
magnitude greater than for SN~1998S at a similar epoch ({\it cf.}
Table 1).  The reason for this is that the $^{56}$Ni mass and the
energy reduction factor, $\psi$, are both $\times3$ lower for SN~1987A
than for SN~1998S, leading to $\sim10$ more efficient CO formation in
SN~1987A.

\section{Conclusions}
We have presented and discussed optical and infrared spectra of the
type~IIn SN~1998S, covering epochs from a few days after explosion to
over 100~days later.  Our observational coverage of SN~1998S makes a
significant contribution to the study of the type~IIn phenomenon.  Of
particular note is our acquisition of (a) contemporary spectra in both
the optical and IR bands at a range of epochs, and (b) the most
extensive set of high resolution spectra ever for this type of
supernova event.  The spectroscopic evolution of SN~1998S was complex.
It can be understood in terms of the interaction of the supernova
with a two-component progenitor wind which we have referred to as the
ICSM and OCSM. 

Collision of the ejecta with the ICSM accounts for the early spectral
features. From the fading of the broad emission components by
$\sim$14d we deduce that the outer boundary of this wind lay at less
than 90~AU from the centre.  Estimation of the inner wind velocity is
difficult due to (a) the ongoing interaction of the shock with this
wind even at the earliest of times, and (b) the lack of high
resolution spectra at these times. We can only say that the early-time
spectra exhibited features of width less than 400~km/s.  However,
these may have arisen in the OCSM.  If, like L00, we assume that the
inner and outer winds had similar velocities then from our high
resolution measurements of the OCSM we can infer that the ICSM wind
commenced less than 9~years ago. However, the possibility of an RSG-BSG
evolution would argue against this assumption.

Examination of the spectra indicates that the OCSM extended from
185~AU to over 1800~AU.  Our high resolution spectra have revealed
that the OCSM has a velocity of 40--50~km/s.  Assuming a constant
velocity during this time, we can infer that it commenced more than
170 years ago, and ceased about 20~years ago.  During this period the
outflow was high - at least $2 \times 10^{-5}$M$_{\odot}$~yr$^{-1}$,
corresponding to a mass loss of at least 0.003~$M_{\odot}$.  An
outflow of this strength and velocity is similar to those seen
in cool supergiants.  The broader absorption feature ($\sim$350~km/s)
in H and He~I may have arisen from a component of the outer CSM shell
produced when the progenitor was going through a later blue supergiant
phase. Alternatively, it may have been due to UV-precursor radiative
acceleration of the inner part of the OCSM.

Our analysis of the CO emission together with the bolometric light
curve Fassia {\it et al.} 2000) also indicates a massive progenitor,
with a mixed metal/He core of $M\sim4~M_{\odot}$ i.e. comparable to
that of SN~1987A.  SN~1998S is only the third core-collapse SN for
which post-100~day $K$-band spectra have been published, and yet all
SN three events have exhibited first-overtone CO emission. (Fassia
{\it et al.} (1998) also published IR spectra for the core-collapse
SN~1995V.  No CO was identified but the latest $K$-band spectrum was
only at 69~days.)  A picture is therefore beginning to emerge where CO
plays a ubiquitous role in the evolution of core-collapse SNe at late
times.  In particular, the powerful cooling property of CO is likely
to lead to conditions in which dust may condense in the ejecta.
Excess IR emission has been reported in SN~1998S on days~225, 260 and
255 (G00), day~253 (Garnavich {\it et al.} 1998) and for a series of
dates to as late as day~691 (Fassia et al. 2000), and this may be due
to dust condensation.  (An IR excess at 130~days was also reported by
Fassia {\it et al.} (2000) but they argue that it was unlikely to have
arisen from dust condensation).  In a future paper (Fassia {\it et al.}
in preparation) we shall examine in detail the evolution of the
late-time IR excess.

L00 used spectropolarimetry at early times to infer significant
asymmetry in the OCSM and in the ejecta-ICSM interface.  While our
discussion of the high resolution OCSM lines assumed spherical
symmetry, we have not provided quantitative explanations for the
relative shifts between the allowed and forbidden lines.  It may be
that this will also require the introduction of an asymmetric model.
G00 also deduced asymmetry in the OCSM, on the basis of the broad line
profiles of H and He on days~240, 275, 370.  We shall discuss these
profiles in a future paper.

\section*{Acknowledgements} 
We thank Janet Drew and Leon Lucy for helpful discussions.  We also
thank Daryl Willmarth for obtaining some of the WIYN data.  The WIYN
Observatory is a joint facility of the University of
Wisconsin-Madison, Indiana University, Yale University, and the
National Optical Astronomy Observatories.  The INT, WHT are operated
on the island of La Palma by the Isaac Newton Group in the Spanish
Observatorio del Roque de los Muchachos of the Instituto de
Astrofisica de Canarias.  We also acknowledge that some of the data
presented here were obtained as part of the UKIRT Service Programme.

\end{document}